\documentclass[aps, prd, amsmath, floats, floatfix, twocolumn, nofootinbib, superscriptaddress]{revtex4-1}
\usepackage[pdftex]{graphicx}
\usepackage{color}
\usepackage{latexsym}
\usepackage[colorlinks]{hyperref}

\newcommand{\be}{\begin{eqnarray}}
\newcommand{\ee}{\end{eqnarray}}

\begin{document}

\title{Relativistic reflection spectra of super-spinning black holes}

\author{Biao~Zhou}
\affiliation{Center for Field Theory and Particle Physics and Department of Physics, Fudan University, 200438 Shanghai, China}

\author{Ashutosh~Tripathi}
\affiliation{Center for Field Theory and Particle Physics and Department of Physics, Fudan University, 200438 Shanghai, China}

\author{Askar~B.~Abdikamalov}
\affiliation{Center for Field Theory and Particle Physics and Department of Physics, Fudan University, 200438 Shanghai, China}

\author{Dimitry~Ayzenberg}
\affiliation{Center for Field Theory and Particle Physics and Department of Physics, Fudan University, 200438 Shanghai, China}

\author{Cosimo~Bambi}
\email[Corresponding author: ]{bambi@fudan.edu.cn}
\affiliation{Center for Field Theory and Particle Physics and Department of Physics, Fudan University, 200438 Shanghai, China}

\author{Sourabh~Nampalliwar}
\affiliation{Theoretical Astrophysics, Eberhard-Karls Universit\"at T\"ubingen, 72076 T\"ubingen, Germany}

\author{Menglei~Zhou}
\affiliation{Center for Field Theory and Particle Physics and Department of Physics, Fudan University, 200438 Shanghai, China}

\begin{abstract}
We construct a relativistic reflection model in a non-Kerr spacetime in which, depending on the value of the deformation parameter of the metric, there are black hole solutions with spin parameter $|a_*| > 1$. We apply our model to fit \textsl{Suzaku} data of four Seyfert galaxies (Ton~S180, Ark~120, 1H0419--577, and Swift~J0501.9--3239). These galaxies host at the center supermassive black holes that were previously interpreted as near-extremal Kerr black holes. For Ton~S180 and 1H0419--577, our measurements are still consistent with the Kerr hypothesis. For Ark~120 and Swift~J0501.9--3239, the Kerr solution is not recovered at 3-$\sigma$. We discuss our results and possible systematic uncertainties in the model.
\end{abstract}

\maketitle

%%%%%%%%%%%%%%%%%%%%%%%%%%%%%%%

\section{Introduction}

In 4-dimensional Einstein's theory of general relativity, uncharged black holes are described by the Kerr solution~\cite{kerr} and are completely characterized by two parameters, which are associated, respectively, to the mass $M$ and the spin angular momentum $J$ of the compact object. Kerr black holes are subject to the constraint $|a_*| \le 1$, where $a_* = J/M^2$ is the dimensionless spin parameter. For $| a_* | > 1$, there is no horizon and the Kerr metric describes the spacetime of a naked singularity.

The spacetime metric around an astrophysical black hole formed from the complete collapse of a progenitor body should be well approximated by the stationary, axisymmetric, and asymptotically-flat Kerr metric, because initial deviations from the Kerr background~\cite{k1}, the presence of nearby stars~\cite{k2}, the gravitational field of accretion disks~\cite{k3}, as well as possible non-vanishing electric charges~\cite{k4} have normally a very small impact on the spacetime metric. All the available black hole observations are consistent with the hypothesis that these objects are the Kerr black holes of Einstein's gravity~\cite{valtonen,ligo,yyp,noi,eht,ghez}. However, there are also a number of theoretical arguments motivating more precise tests that could potentially discover new physics~\cite{m1,m2,n1,n2,n3,n4}.

The spacetime metric around a black hole can be tested with electromagnetic and gravitational wave techniques. The two methods are complementary, as they can probe different sectors of the theory. Electromagnetic tests~\cite{em1,em2,em3,em4,em5,em6,em7,em7b,em8,em8b,em8c,em9,em10,em11} are usually more suitable to explore the interactions between the matter and the gravity sectors, including violation of Einstein's Equivalence Principle~\cite{will}. Gravitational wave tests~\cite{gw1,gw2,gw3,gw4,gw5,gw6} can more directly verify the validity of the Einstein Equations~\cite{barausse}.

If we want to test the nature of astrophysical black holes with electromagnetic techniques, the most logical method would be compare the predictions of general relativity with those of another theory of gravity in which uncharged black holes are not described by the Kerr solution\footnote{In the case of gravitational wave tests, two theories may predict different signals even in the case the background metric is the same, because the field equations are different~\cite{barausse}.}. In such a case, we should analyze the observational data of a specific source with the Kerr model of general relativity and with the non-Kerr model of the other theory of gravity, and then check if observations prefer one of the two models and can rule out the other one. This is the so-called top-down approach. However, there are two problems. First, there are many theories of gravity, so we should repeat the same test for a large number of theories. Second, usually we do not know the rotating black hole solutions in theories beyond general relativity! This is just a technical problem to solve the corresponding field equations: while it is relatively easy to find spherically symmetric, non-rotating black hole solutions, it is definitively more difficult to find the complete axisymmetric, rotating black hole metrics of a theory of gravity.

The alternative approach, which is often preferred in literature, is the so-called bottom-up method. Now we try to consider an extension of the Kerr metric by adding extra pieces to the Kerr solution. The magnitude of these extra terms is regulated by some ``deformation parameters'', which are used to quantify possible deviations from the Kerr background. These spacetimes are commonly called parametric black hole spacetimes because they are black hole metrics in which deviations from the Kerr background are parametrized by the deformation parameters. With the spirit of performing a null experiment, we analyze astronomical data with a model assuming such an extended Kerr metric and we try to constrain the value of the deformation parameters to check whether observations require that these parameters vanish, thus confirming the Kerr solution of Einstein's gravity.

Among the electromagnetic methods, X-ray reflection spectroscopy is quite a promising tool to test astrophysical black holes~\cite{r1,r2,r3}. Accreting black holes can be surrounded by geometrically thin and optically thick disks when the accreting gas has a large angular momentum and the accretion luminosity is between a few percent to about 30\% of the Eddington limit of the source. The temperature of the inner part of these accretion disks is in the soft X-ray band for stellar-mass black holes ($M \sim 10$~$M_\odot$) and in the optical/UV band for supermassive black holes ($M \sim 10^5$-$10^{10}$~$M_\odot$). Thermal photons from the disk can inverse Compton scatter off free electrons of the so-called corona, which is a generic name to indicate a hotter ($\sim 100$~keV) cloud of gas near the black hole. The corona may be represented by the accretion flow plunging from the disk to the black hole, the base of the jet, etc. Recent studies of the X-ray reverberation lags in both black hole binaries and Seyfert AGNs suggest a compact coronal region, even if the geometry cannot be precisely measured~\cite{Kara:2014jda,Kara:2019zad}. The process of inverse Compton scattering generates a power-law component with an exponential cut-off correlated to the corona temperature. A fraction of this power-law component can illuminate the disk, producing a reflection component. X-ray reflection spectroscopy refers to the analysis of this reflection component.

In the rest-frame of the accreting gas, the reflection spectrum is characterized by some fluorescent emission lines in the soft X-ray band and by the Compton hump at 20-30~keV. The most prominent emission feature is often the iron K$\alpha$ complex, which is at 6.4~keV in the case of neutral or weakly ionized iron and shifts up to 6.97~keV in the case of H-like iron ions (while there is no line in the case of fully ionized iron). The reflection spectrum of the disk far from the source is the result of the combination of the reflection components from different points of the disk, each of them differently shifted by relativistic effects occurring in the strong gravity region near the black hole. For a given accretion disk model and a given spacetime metric, we can compute the reflection spectrum of the whole disk far from the source from the reflection spectrum at the emission point. In the reality, we can construct a model with a reflection spectrum at the emission point, an accretion disk, and a metric, and we can fit the data of an accreting black hole to estimate the values of the model parameters.

In Refs.~\cite{v1,v2}, we have presented {\sc relxill\_nk}, which is an extension of the {\sc relxill} package~\cite{j1,j2} to parameterise black hole spacetimes. In our previous studies, we have mainly considered the Johannsen metric~\cite{johannsen}, but the model can be easily implemented for any stationary, axisymmetric, and asymptotically-flat black hole spacetime. From the analysis of X-ray data of astrophysical black holes with {\sc relxill\_nk}, we can measure the deformation parameters of the spacetime and thus constrain possible deviations from the Kerr geometry.

Black holes that were previously found to be spinning at a near-maximum spin are the most suitable sources for this kind of tests, because the inner edge of the accretion disk can be very close to the compact object, relativistic features in the reflection spectrum are maximized, and it is thus possible to break the parameter degeneracy and get good measurements of the spacetime metric. For some supermassive black holes, we have obtained remarkably strong constraints on possible deviations from the Kerr metric and spin parameters stuck to the maximum value allowed by the model; see Refs.~\cite{bare,gs1354} for more details. However, such extreme values of the spin parameter have to be taken with caution. Best fit values at the boundary of the parameter space allowed by the model could indeed be a sympton of the breakdown of the model itself.

In the present work, we further explore the opportunities to test the Kerr hypothesis with such apparently near-extremal Kerr black holes. We consider a subset of the black hole spacetimes proposed in Ref.~\cite{nan}. In addition to the mass and the spin angular momentum, these spacetimes are characterized by the deformation parameter $\alpha$. For $\alpha = 0$, we exactly recover the Kerr solution. For $\alpha > 0$, there are solutions describing black holes with $|a_*| > 1$ and very small values of the radial coordinate of the event horizon. Implementing such a metric in {\sc relxill\_nk}, we fit \textsl{Suzaku} observations of four supermassive black holes (Ton~S180, Ark~120, 1H0419--577, and Swift~J0501.9--3239) that have been previously interpreted as near-extremal Kerr black holes~\cite{bare,w13,Jiang:2018aln,Jiang:2019ztr,Agis-Gonzalez:2014nja}.

The paper is organized as follows. In Section~\ref{s-metric}, we review the black hole metric proposed in Ref.~\cite{nan}. In Section~\ref{s-xrs}, we construct a relativistic reflection model for such a metric. In Section~\ref{s-sources}, we apply our model to analyze \textsl{Suzaku} data of four supermassive black holes: Ton~S180, Ark~120, 1H0419--577, and Swift~J0501.9--3239. Our results are discussed in Section~\ref{s-dc}. Throughout the paper, we employ units in which $G_{\rm N} = c = 1$ and the convention of a metric with signature $(-+++)$.

%%%%%%%%%%%%%%%%%%%%%%%%%%%%%%%

\section{Super-spinning black holes \label{s-metric}}

In this work, we want to consider a subset of the family of black hole metrics proposed in Ref.~\cite{nan}. In Boyer-Lindquist coordinates, the line element reads
\be\label{eq-ds}
ds^2 &=& - \left( 1 - \frac{2 m_1 r}{\Sigma} \right) dt^2 
+ \frac{\Sigma}{\Delta} dr^2 + \Sigma d\theta^2 \nonumber\\
&& 
+ \left( r^2 + a^2 + \frac{2 a^2 m_1 r \sin^2\theta}{\Sigma} \right) \sin^2\theta \, d\phi^2
 \nonumber\\
&& - \frac{4 a m_1 r \sin^2\theta}{\Sigma} dt \, d\phi \, ,
\ee
where
\be
\Sigma &=& r^2 + a^2 \cos^2\theta \, , \\
\Delta &=& r^2 - 2 m_2 r + a^2 \, ,
\ee
and $a = J/M$. Eq.~(\ref{eq-ds}) reduces to the Kerr solution for $m_1 = m_2 = M$. In general, $m_1 = m_1 (r)$ and $m_2 = m_2 (r)$ are functions of the radial coordinate $r$ (this gives the metric some nice properties, including the separability of the equations of motion~\cite{nan}). In what follows, we will only consider the special case
\be
m_1 = m_2 = M \left( 1 + \alpha \frac{M^2}{r^2} \right) \, .
\ee
$\alpha$ is the deformation parameter of the metric and the Kerr solution is recovered for $\alpha = 0$. In general, we could write $m_1$ and $m_2$ as an expansion in $M/r$. However, the first order term can be constrained by Solar System experiments. Our form of $m_1$ and $m_2$ is thus the simplest choice without constraints from tests of general relativity in the weak field regime~\cite{nan}.

The metric in Eq.~(\ref{eq-ds}) has some interesting properties. In particular, it has very compact and fast-rotating black holes, and matter falling onto similar objects can release more energy than matter falling onto a Kerr black hole~\cite{nan}. This is quite a relevant property because: $i)$ extremal Kerr black holes are characterized by a very high radiative efficiency and, when we consider deformations of the Kerr metric, we usually obtain black holes with lower radiative efficiency, and $ii)$ observations show that astrophysical black holes can have a very high radiative efficiency, which often helps to get strong constraints on possible deviations from the Kerr solution.

\begin{figure*}[t]
\begin{center}
\includegraphics[width=8.5cm]{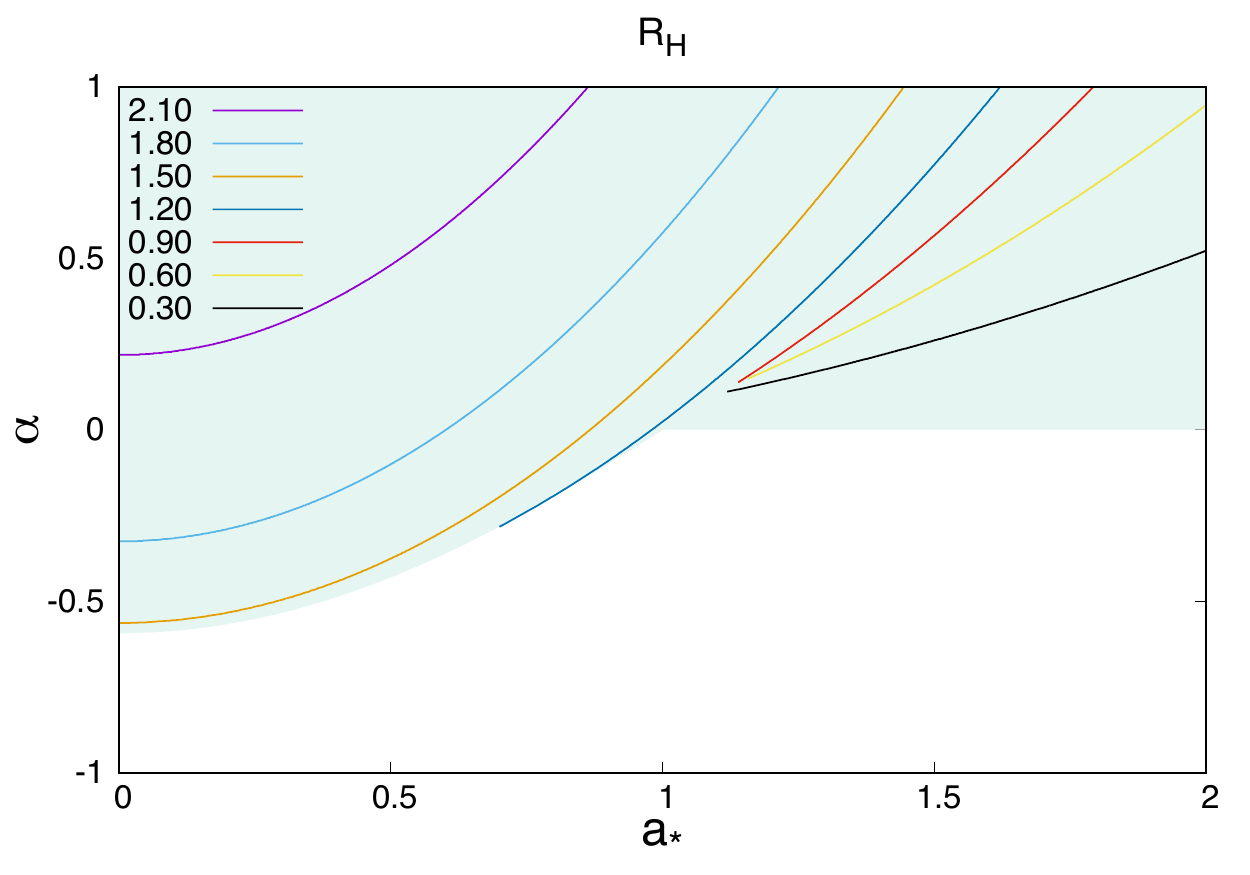}
\includegraphics[width=8.5cm]{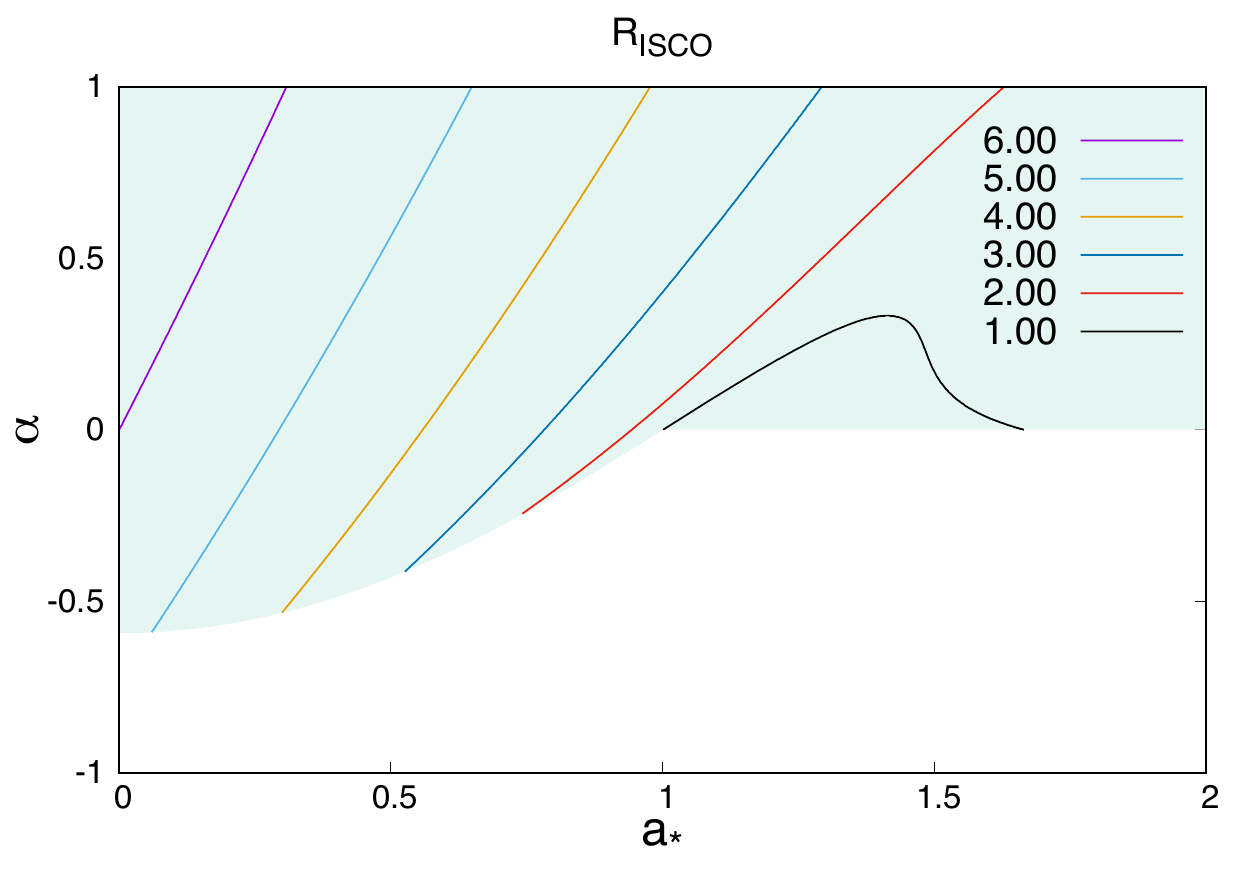}
\end{center}
\vspace{-0.5cm}
\caption{Contour levels of the radial coordinates of the event horizon $R_{\rm H}$ (left panel) and of the ISCO $R_{\rm ISCO}$ (right panel) in the plane $a_*$ vs $\alpha$ in units $M=1$. The parameter space in white is ignored in our study because the spacetimes there have no black holes but naked singularities. \label{f-metric}}
\end{figure*}

The radial coordinate of the event horizon is given by the largest root of $\Delta = 0$. For $\alpha = 0$, we recover the Kerr result $R_{\rm H} = M + \sqrt{M^2 - a^2}$, which ranges from $2M$ for a non-rotating black hole to $M$ for an extremal black hole with $|a_*| = 1$, and there is no horizon for $|a_*| > 1$. For arbitrary $\alpha$, we find a richer phenomenology. As shown in the left panel of Fig.~\ref{f-metric}, for $\alpha > 0$ $R_{\rm H}$ can be either larger than $2M$ or smaller than $M$, depending on the exact values of $a_*$ and $\alpha$. Moreover, the horizon still exists for $|a_*| > 1$. The boundary between spacetimes with black holes and spacetimes with naked singularity is described by
\be
\alpha = \left\{ \begin{array}{ll}
\frac{1}{2} \left( x^3 - 2 x^2 + a^2_* x \right) & \text{ for $|a_*| \le 1$} \\
0 & \text{ for $|a_*| > 1$}
\end{array} \right.
\ee
where
\be
x = \frac{2}{3} + \sqrt{\frac{4}{9} - \frac{a_*^2}{3}} \, .
\ee

Another important quantity is the radial coordinate of the innermost stable circular orbit (ISCO), $R_{\rm ISCO}$. In the Kerr metric, it monotonically decreases from $9M$ for an extremal black hole and a counterrotating orbit to $6M$ for a non-rotating black hole and to $M$ for an extremal black hole and a corotating orbit. As shown in the right panel of Fig.~\ref{f-metric}, for $\alpha > 0$ the ISCO radius can be smaller than $M$.

%%%%%%%%%%%%%%%%%%%%%%%%%%%%%%%

\section{X-ray reflection spectroscopy \label{s-xrs}}

In Refs.~\cite{v1,v2}, we have presented the XSPEC-compatible, public, relativistic reflection model {\sc relxill\_nk}, which is the natural extension of the {\sc relxill} package~\cite{j1,j2} to non-Kerr spacetimes. The model employs the approach of the transfer function proposed by Cunningham~\cite{trf}. The observed flux (measured, for instance, in erg~cm$^{-2}$~s$^{-1}$~Hz$^{-1}$) of the reflection spectrum is obtained by performing the integral
\be\label{eq-flux}
F_{\rm o} (\nu_{\rm o}) = \frac{1}{D^2} \int_{r_{\rm in}}^{r_{\rm out}} \int_0^1 \frac{\pi r_{\rm e} g^2}{\sqrt{g^* \left(1 - g^* \right)}} \, f \, I_{\rm e} \, dg^* \, dr_{\rm e} \, , \,\,\,\,\,\,
\ee
where $D$ is the distance of the source from the detection point, $r_{\rm in}$ and $r_{\rm out}$ are, respectively, the inner and the outer edge of the accretion disk, $g = \nu_{\rm o}/\nu_{\rm e}$ is the redshift factor, $\nu_{\rm o}$ and $\nu_{\rm e}$ are, respectively, the photon frequencies at the detection point far from the source and at the emission point in the rest-frame of the accreting gas, $r_{\rm e}$ is the radial coordinate of the emission point, and $I_{\rm e}$ is the specific intensity of the radiation at the emission point in the rest-frame of the accreting gas. $g^*$ is the relative redshift factor, which is defined as
\be
g^* = \frac{g - g_{\rm min}}{g_{\rm max} - g_{\rm min}}
\ee
and ranges from 0 to 1. $g_{\rm max} = g_{\rm max} (r_{\rm e}, i)$ and $g_{\rm min} = g_{\rm min} (r_{\rm e}, i) $ are, respectively, the maximum and the minimum redshift factor for the photons emitted at the radial coordinate $r_{\rm e}$ in the accretion disk and when the viewing angle is $i$. $f = f (g^*, r_{\rm e}, i)$ is the transfer function defined as
\be
f = \frac{g \sqrt{g^* \left(1 - g^*\right)}}{\pi r_{\rm e}} \, \left| \frac{\partial \left(X,Y\right)}{\partial \left(g^*,r_{\rm e}\right)} \right| \, ,
\ee
and $| \partial \left(X,Y\right) / \partial \left(g^*,r_{\rm e}\right) |$ is the Jacobian between the Cartesian coordinates $X$ and $Y$ of the image of the disk in the plane of the distant observer and the disk variables $g^*$ and $r_{\rm e}$. More details can be found in Refs.~\cite{trf,v1,v2}

In Eq.~(\ref{eq-flux}), the relativistic effects affecting the reflection spectrum are encodes in the transfer function $f$ (and, at some level, in the radial profile of $I_{\rm e}$ if we consider a specific coronal geometry, which is not our case here). Since the evaluation of the transfer function requires the calculation of a large number of null geodesics, from the plane of the distant observer backwards in time to the emission point on the accretion disk, and these calculations are quite time consuming, it is convenient to tabulate the transfer functions into a FITS (Flexible Image Transport System) file. {\sc relxill\_nk} reads the FITS file and calculate the integral in Eq.~(\ref{eq-flux}) during the data analysis.

\begin{figure*}[t]
\begin{center}
\includegraphics[width=8.5cm]{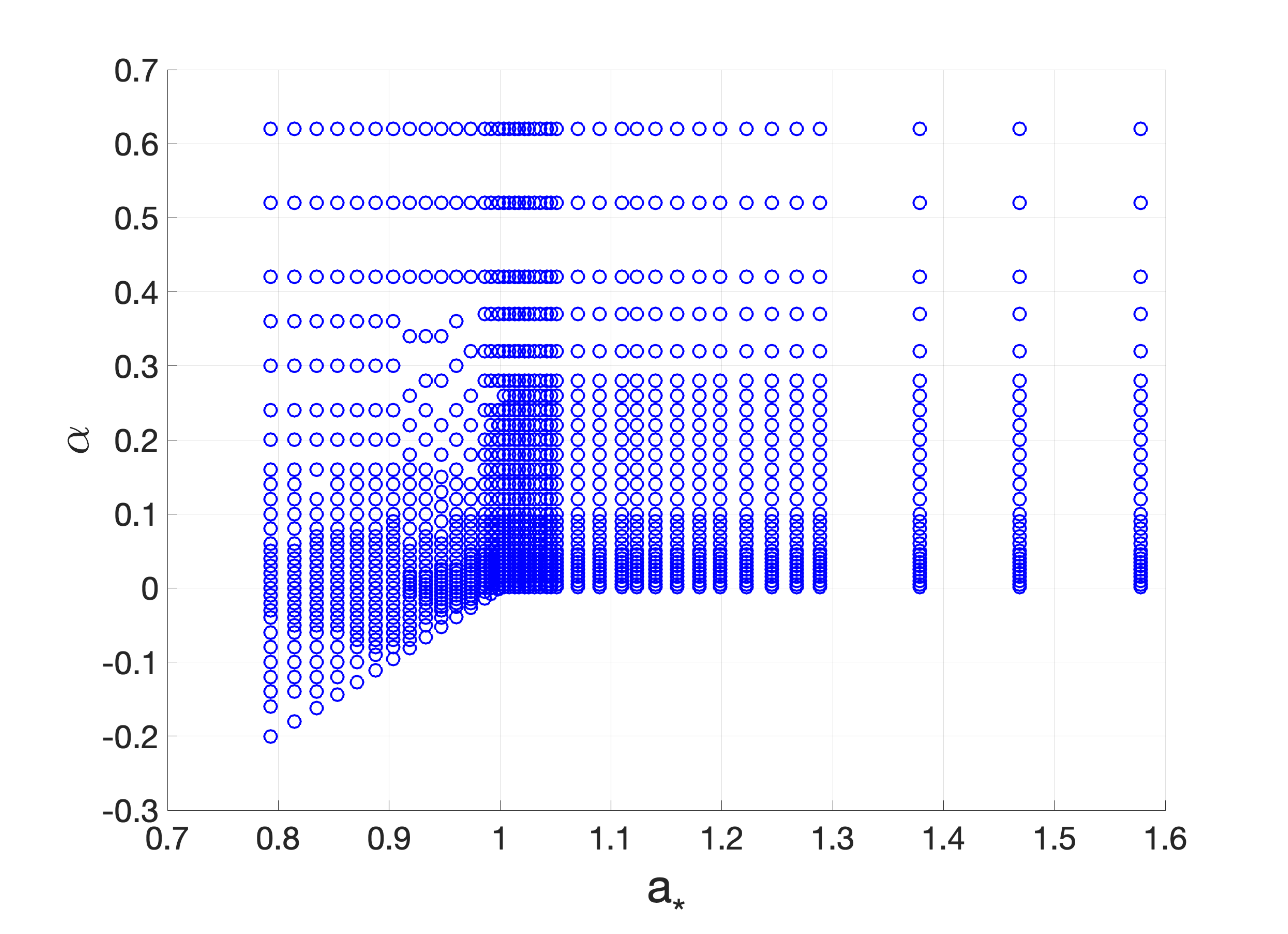}
\includegraphics[width=8.5cm]{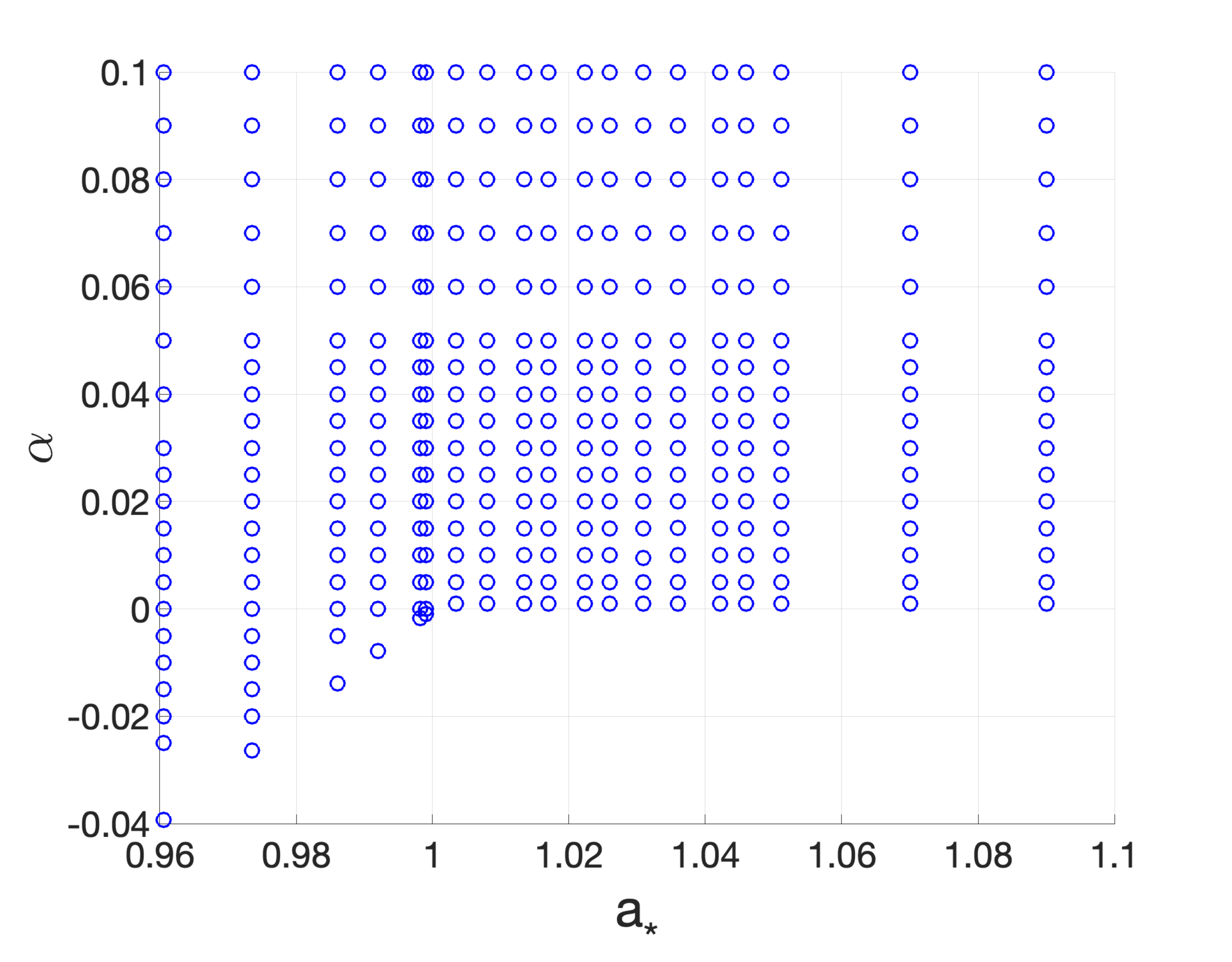}
\end{center}
\vspace{-0.4cm}
\caption{{The left panel shows the grid points in the FITS file for the spin parameter $a_*$ and the deformation parameter $\alpha$. The right panel zooms in on the region near $a_* = 1$ and $\alpha = 0$.} \label{f-grid}}
\end{figure*}

In the FITS file, every transfer function is specified by its values of the black hole spin parameter $a_*$, deformation parameter $\alpha$, and inclination angle of the disk $i$, which form a grid $42 \times 30 \times 22$. Transfer functions of generic configurations are obtained by interpolation of the transfer functions of this grid. The grid points are more dense where the ISCO radius changes more rapidly. Fig.~\ref{f-grid} shows the distribution of the grid points in the parameter space $a_*$ vs $\alpha$. The grid points for the inclination angle of the disk are evenly distributed in $0 < \cos i < 1$, as discussed in Ref.~\cite{v2}.

%%%%%%%%%%%%%%%%%%%%%%%%%%%%%%%

\section{Data analysis \label{s-sources}}

In this section, we apply our reflection model with the metric in Eq.~(\ref{eq-ds}) to the \textsl{Suzaku} data of four supermassive black holes: Ton~S180, Ark~120, 1H0419--577, and Swift~J0501.9--3239. The choice of these sources is not accidental, but determined by some remarkable properties. First, these objects were analyzed in Ref.~\cite{bare}, where it was found that they are compatible with near-extremal Kerr black holes. The best-fit value of their spin parameter was found to be 0.998, or close to it, which was the maximum value allowed by the model. Deviations from the Kerr metric, if any, were constrained to be small. Moreover, these four sources present quite a simple spectrum, with no intrinsic absorption complicating the data analysis. Further details can be found in Ref.~\cite{bare} as well as in Ref.~\cite{w13}, where these data were analyzed for the first time.

\begin{table}[b]
\centering
\vspace{0.3cm}
\begin{tabular}{lccc}
\hline\hline
Source & Observation ID & Year & Exposure (ks) \\ 
\hline\hline
Ton~S180 & 701021010 & 2006 & 108 \\
\hline
Ark~120 & 702014010 & 2007 & 91 \\
\hline
1H0419--577 & 702041010 & 2007 &179 \\
\hline
Swift~J0501.9--3239 & 703014010 & 2008 & 36 \\
\hline\hline
\end{tabular}
\vspace{0.1cm}
\caption{Sources and \textsl{Suzaku} observations analyzed in our study.}
\label{t-obs}
\end{table}

The basic details of the \textsl{Suzaku} observations analyzed in our study are reported in Tab.~\ref{t-obs}. The reduction of these data was already discussed in Ref.~\cite{bare}. In summary, we use the data of the three front-illuminated CCD detectors (XIS0, XIS2 if available\footnote{XIS2 experienced charge leakage on 6~November~2007 and therefore there are no XIS2 data for observations after that date. This is the case only of Swift~J0501.9--3239 in the present study.}, XIS3). The back-illuminated CCD detector has a smaller effective area around 6~keV and a higher background at higher energies, so the inclusion/exclusion of its data does not appreciably affect our fits. Similarly, we do not include HXD PIN data in our study because their quality is too poor and do not permit to improve the fits. We use {\sc heasoft} v6.24 and CALDB version 20180312 for data reduction. Spectra and response files are combined using the {\sc ftool} ADDASCASPEC. The combined spectra are rebinned to a minimum of 50~counts in order to ensure the validity of the $\chi^2$ fit statistics. In the data analysis, we ignore the energy range 1.7-2.5~keV because of calibration uncertainties\footnote{See https://heasarc.gsfc.nasa.gov/docs/suzaku/analysis/sical.html for more details.}.

\begin{figure*}[t]
\begin{center}
\includegraphics[width=8.5cm,trim={0.5cm 0 3cm 18.3cm},clip]{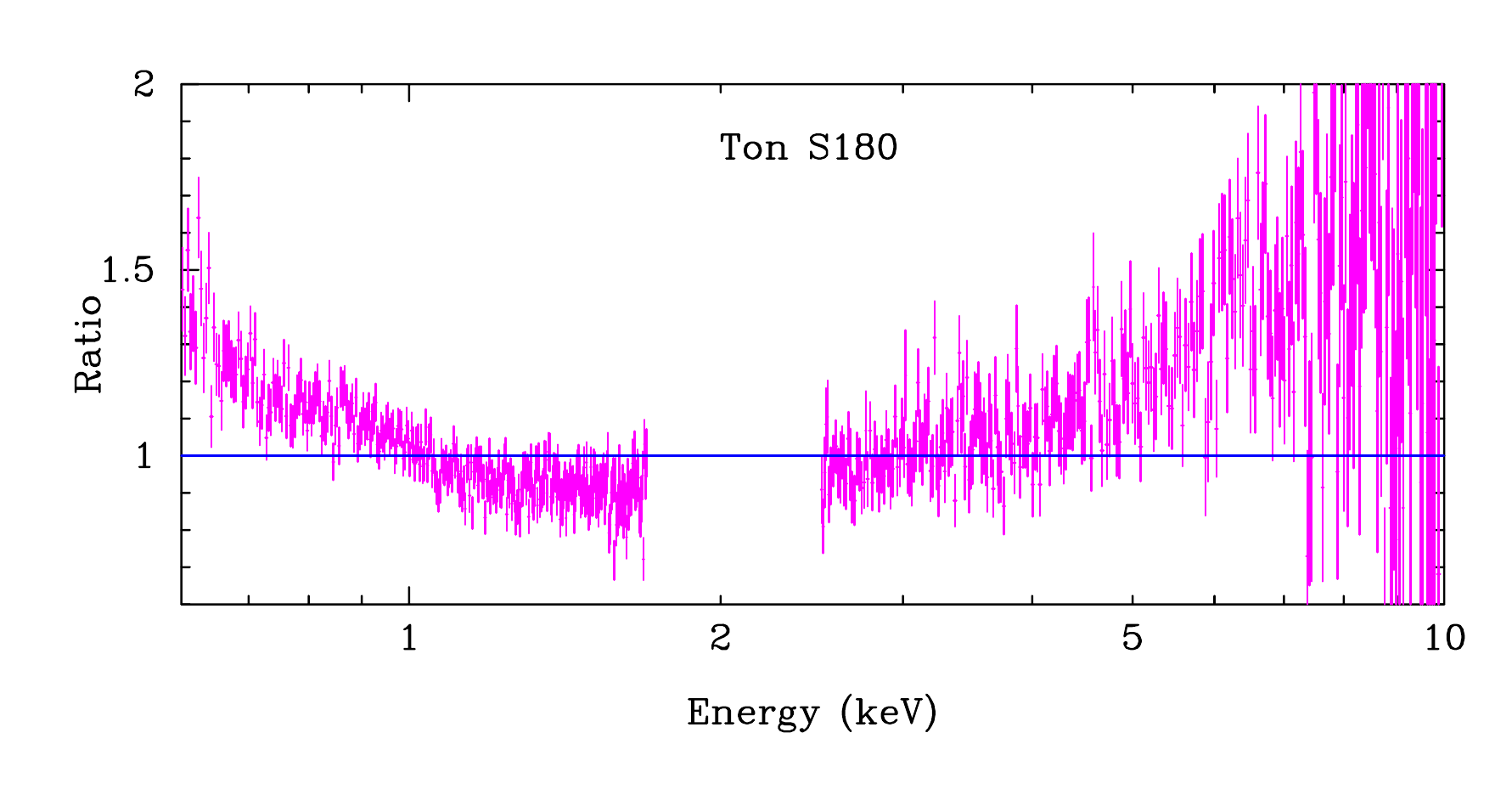}
\includegraphics[width=8.5cm,trim={0.5cm 0 3cm 18.3cm},clip]{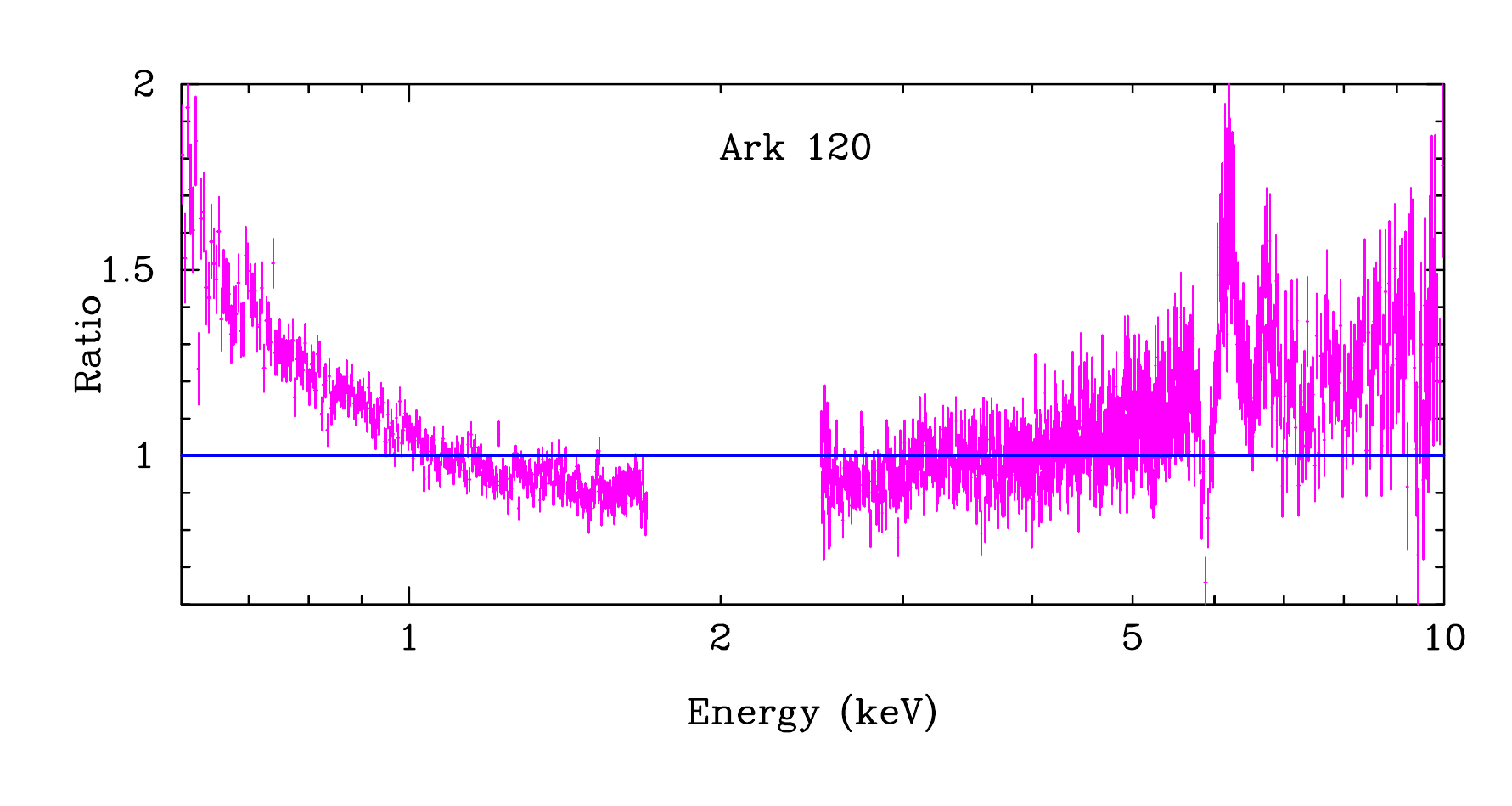} \\
\includegraphics[width=8.5cm,trim={0.5cm 0 3cm 18.3cm},clip]{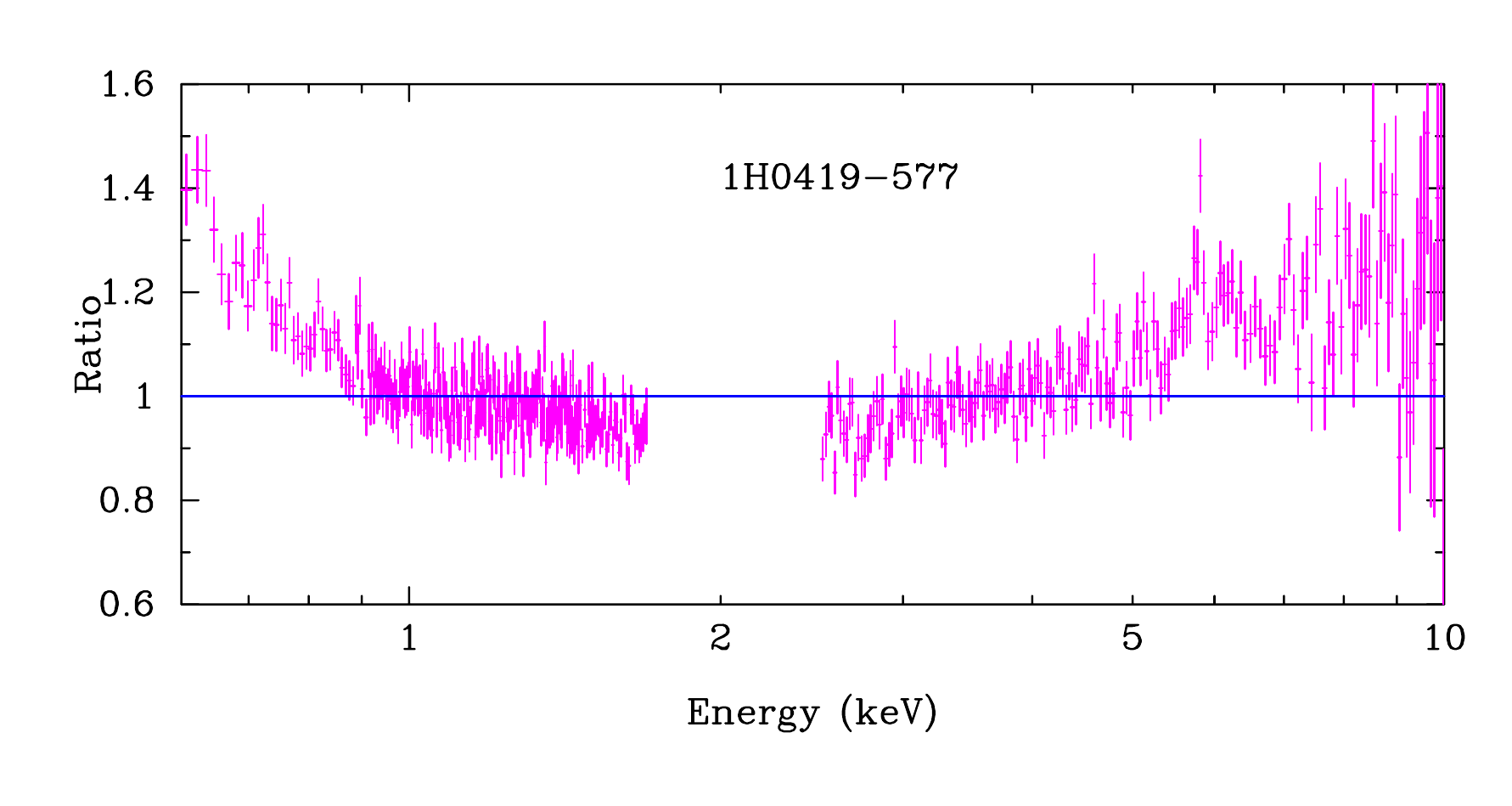}
\includegraphics[width=8.5cm,trim={0.5cm 0 3cm 18.3cm},clip]{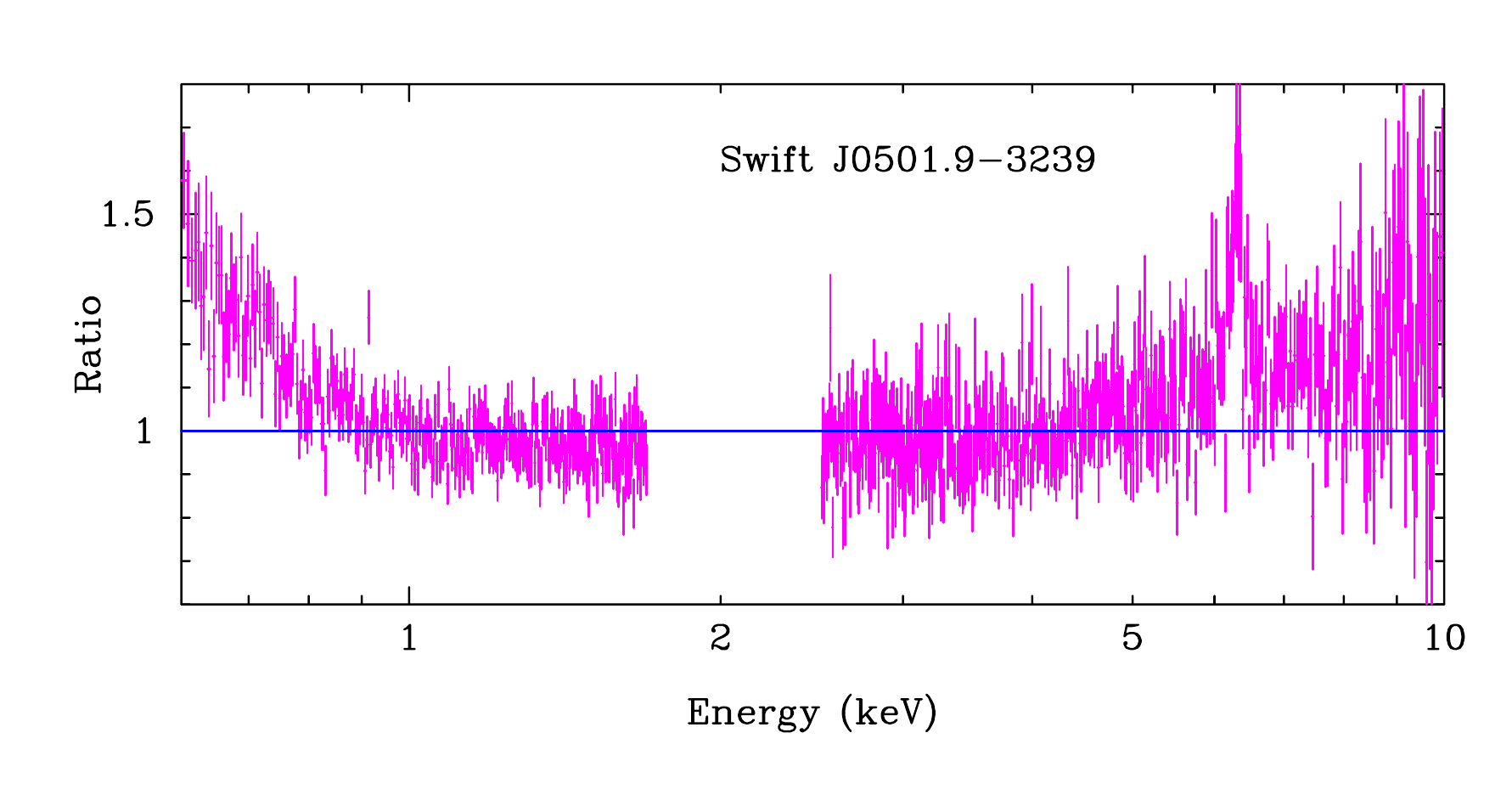}
\end{center}
\vspace{-0.8cm}
\caption{{Data-to-best-fit-model ratios for the four sources of our study when the spectra are described by a power-law component only.} \label{f-p}}
\end{figure*}

Fig.~\ref{f-p} shows the data-to-best-fit-model ratios for the four sources when the model is described by an absorbed power-law. All spectra are characterized by a strong soft excess below 2~keV and a relatively broad iron line at 5-7~keV. The interpretation of the soft excess as a relativistic reflection feature is also crucial for the determination of the model parameters in the analysis of the present work.

The data analysis follows Ref.~\cite{bare}. We use XSPEC v12.9.1~\cite{arnaud}. For Ton~S180, the XSPEC model is 

\vspace{0.2cm}

{\sc tbabs$\times$(zpowerlw + relxill\_nk)} .

\vspace{0.2cm}

\noindent For Ark~120, the XSPEC model is

\vspace{0.2cm}

{\sc tbabs$\times$(zpowerlw + relxill\_nk + xillver + zgauss + zgauss)} .

\vspace{0.2cm}

\noindent Lastly, the data of 1H0419--577 and Swift~J0501.9--3239 are fitted with the XSPEC model

\vspace{0.2cm}

{\sc tbabs$\times$(zpowerlw + relxill\_nk + xillver)} .

\vspace{0.2cm}

{\sc tbabs} describes the Galactic absorption~\cite{wilms}. The hydrogen column density, $n_{\rm H}$, is not a free parameter in our fit and is frozen to the value calculated using the tool at\footnote{http://www.swift.ac.uk/analysis/nhtot/}~\cite{Willingale:2013tia}.

{\sc zpowerlw} describes the power-law component from the corona. The photon index $\Gamma$ and the normalization of the component are free parameters in the fit, while the redshift of the source is frozen to the value reported on the NASA extragalactic database (NED); see Tab.~\ref{t-fit} for the input value of every source.

{\sc relxill\_nk} describes the reflection spectrum from the disk~\cite{v1,v2}. The inclination angle of the disk $i$, the spin parameter $a_*$, the ionization parameter $\xi$, and the iron abundance $A_{\rm Fe}$ are always free parameters in the fits. For every source, first we fit the data assuming the Kerr metric (so imposing $a_* \le 0.998$ and $\alpha = 0$) and then for the general case with both $a_*$ and $\alpha$ completely free. For the intensity profile, we consider either a simple power-law or a broken power-law, and we choose the model that provides the best result. Eventually, for Ton~S180 and Ark~120, we employ a broken power-law with outer emissivity index frozen to 3 (lamppost coronal set-up) and therefore we have two free parameters, namely the inner emissivity index $q_{\rm in}$ and the breaking radius $R_{\rm br}$. For 1H0419--577 and Swift~J0501.9--3239, we find that a simple power-law is enough to fit the data, so we have only one free parameter in the fit\footnote{{If we employ a broken power-law for 1H0419--577 and Swift~J0501.9--3239, the quality of the fits improve marginally and we find very similar best-fit values of the model parameters. More specifically, when $\alpha$ is free, for 1H0419--577 the difference of $\chi^2$ is 17. For Swift~J0501.9--3239, we find $\Delta \chi^2 \lesssim 1$ (if both indices are free, we cannot constrain the outer emissivity index and the breaking radius; if we impose $q_{\rm out} = 3$, we recover a large breaking radius and a fit equivalent to the case of a simple power-law). Residuals in the ratio plots are also similar between the models with simple power-law and broken power-law.}}.
The photon index of the radiation illuminating the disk is tied to the value of the photon index in {\sc zpowerlw} and the cut-off energy $E_{\rm cut}$ is frozen to 300~keV because it cannot be measured with the XIS data reaching 10~keV.

With the exception of Ton~S180, the spectra of these sources also show a non-relativistic reflection component from some cold material far from the strong gravity region. Such a component is fitted with {\sc xillver}~\cite{xillver}, where the values of the parameters are tied to those in {\sc relxill\_nk}, with the exception of the ionization parameter, which is frozen to $\log\xi = 0$.

In the data of Ark~120, we also see an emission feature around 6.95~keV (consistent with Fe~XXVI) and an absorption feature around 6.1~keV. We fit these features by adding two narrow lines with {\sc zgauss}.

\begin{table*}
\centering
\vspace{0.3cm}
\begin{tabular}{lcc|cc|cc|cc}
\hline\hline
& \multicolumn{2}{c}{Ton~S180} & \multicolumn{2}{c}{Ark~120} & \multicolumn{2}{c}{1H0419--577} & \multicolumn{2}{c}{Swift~J0501.9--3239} \\
\hline
{\sc tbabs} &&&&&&&& \\
$n_{\rm H}$~[$10^{21}$~cm$^{-2}$] & $0.136^*$ & $0.136^*$ & $1.45^*$ & $1.45^*$ & $0.134^*$ & $0.134^*$ & $0.184^*$ & $0.184^*$ \\
\hline
{\sc zpowerlw} &&&&&&&& \\
$\Gamma$ & $2.44_{-0.07}^{+0.04}$ & $2.45_{-0.05}^{+0.04}$ & $2.42_{-0.03}^{+0.07}$ & $2.48_{-0.07}^{+0.11}$ & $2.15_{-0.03}^{+0.04}$ & $2.38_{-0.05}^{+0.03}$ & $2.317_{-0.074}^{+0.019}$ & $2.366_{-0.055}^{+0.013}$ \\
$z$ & $0.062^*$ & $0.062^*$ & $0.0327^*$ & $0.0327^*$ & $0.104^*$ & $0.104^*$ & $0.0124^*$ & $0.0124^*$ \\
\hline
{\sc relxill\_nk} &&&&&&&& \\
$q_{\rm in}$ & $> 9.75$ & $9.9_{-0.6}^{\rm (P)}$ & $9.2_{-0.9}^{+0.7}$ & $8.0_{-0.3}^{+0.6}$ & $7.5_{-1.3}^{+2.0}$ & $7.6_{-1.8}^{+0.6}$ & $> 9.86$ & $9.64_{-0.42}^{+0.10}$ \\
$q_{\rm out}$ & $3^*$ & $3^*$ & $3^*$ & $3^*$ & $= q_{\rm in}$ & $= q_{\rm in}$ & $= q_{\rm in}$ & $= q_{\rm in}$ \\
$R_{\rm br}$ $[M]$ & $3.16_{-0.09}^{+0.17}$ & $3.24_{-0.11}^{+0.24}$ & $4.4_{-0.4}^{+0.7}$ & $4.6_{-0.6}^{+1.2}$ & -- & -- & -- & -- \\
$i$ [deg] & $37_{-3}^{+3}$ & $37.5_{-1.6}^{+2.4}$ & $23_{-6}^{+3}$ & $17_{-7}^{+3}$ & $71_{-3}^{+5}$ & $54_{-9}^{+13}$ & $3_{\rm (P)}^{+4}$ & $13.3_{-2.2}^{+2.1}$ \\
$a_*$ & $0.996_{-0.003}^{\rm (P)}$ & $1.0018_{-0.0018}^{+0.0013}$ & $> 0.9971$ & $1.242_{-0.018}^{+0.022}$ & $> 0.995$ & $1.029_{-0.011}^{+0.020}$ & $0.9945_{-0.0015}^{+0.0009}$ & $1.131_{-0.036}^{+0.019}$ \\
$\alpha$ & $0^*$ & $0.004_{-0.004}^{+0.004}$ & $0^*$ & $0.213_{-0.011}^{+0.030}$ & $0^*$ & $0.02_{-0.02}^{+0.03}$ & $0^*$ & $0.137_{-0.003}^{+0.003}$ \\
$\log\xi$ & $3.27_{-0.06}^{+0.04}$ & $3.305_{+0.195}^{+0.013}$ & $2.999_{+0.214}^{+0.025}$ & $3.07_{+0.32}^{+0.04}$ & $0.69_{-0.27}^{+0.15}$ & $1.55_{-0.23}^{+0.35}$ & $2.85_{-0.12}^{+0.36}$ & $2.87_{-0.19}^{+0.13}$ \\
$A_{\rm Fe}$ & $3.2_{-1.0}^{+0.5}$ & $4.2_{-1.6}^{+2.6}$ & $1.5_{-0.5}^{+0.6}$ & $3.8_{-0.6}^{+2.0}$ & $2.0_{-0.5}^{+0.5}$ & $< 0.75$ & $1.9_{-0.9}^{+0.3}$ & $3.35_{-0.20}^{+0.88}$ \\
$E_{\rm cut}$ [keV] & $300^*$ & $300^*$ & $300^*$ & $300^*$ & $300^*$ & $300^*$ & $300^*$ & $300^*$ \\
\hline
{\sc xillver} &&&&&&&& \\
$\log\xi$ & -- & -- & $0^*$ & $0^*$ & $0^*$ & $0^*$ & $0^*$ & $0^*$ \\
\hline
{\sc zgauss} &&&&&&&& \\
$E_{\rm line}$ [keV] & -- & -- & $6.95_{-0.03}^{+0.03}$ & $6.95_{-0.03}^{+0.03}$ & -- & -- & -- & -- \\
\hline
{\sc zgauss} &&&&&&&& \\
$E_{\rm line}$ [keV] & -- & -- & $6.087_{-0.016}^{+0.015}$ & $6.088_{-0.016}^{+0.011}$ & -- & -- & -- & -- \\
\hline
$\chi^2/\nu$ & 1352.59/1314 & 1350.38/1313 & 1408.69/1309 & 1388.29/1308 & 2488.20/2345 & 2487.41/2344 & 1353.04/1314 & 1346.18/1313 \\
& =1.02937 & =1.02847 & =1.07616 & =1.06138 & =1.06107 & =1.06118 & =1.02971 & =1.02527 \\
$\Delta\chi^2$ & \multicolumn{2}{c}{2.21} & \multicolumn{2}{c}{20.40} & \multicolumn{2}{c}{0.79} & \multicolumn{2}{c}{6.86} \\
\hline\hline
\end{tabular}
%\vspace{0.1cm}
\caption{Best-fit values of the four sources considered in our study. For every source, the left column is for the Kerr model ($a_* \le 0.998$, $\alpha = 0$) and the right column is when $\alpha$ is free in the fit. $^*$ indicates that the parameter is frozen in the fit. All uncertainties are at 90\% confidence for one relevant parameter ($\Delta\chi^2 = 2.71$). (P) indicates that we do not find the 90\% confidence limit in the fit because we reach the bound imposed on the parameter. $\xi$ in units erg~cm~s$^{-1}$. $A_{\rm Fe}$ in units of Solar abundance. In the last row, $\Delta\chi^2$ is the difference between the minimum of $\chi^2$ of the Kerr model and of the model with free $\alpha$.}
\label{t-fit}
\end{table*}

\begin{figure*}[t]
\begin{center}
\includegraphics[width=8.5cm,trim={0.5cm 0 3cm 18cm},clip]{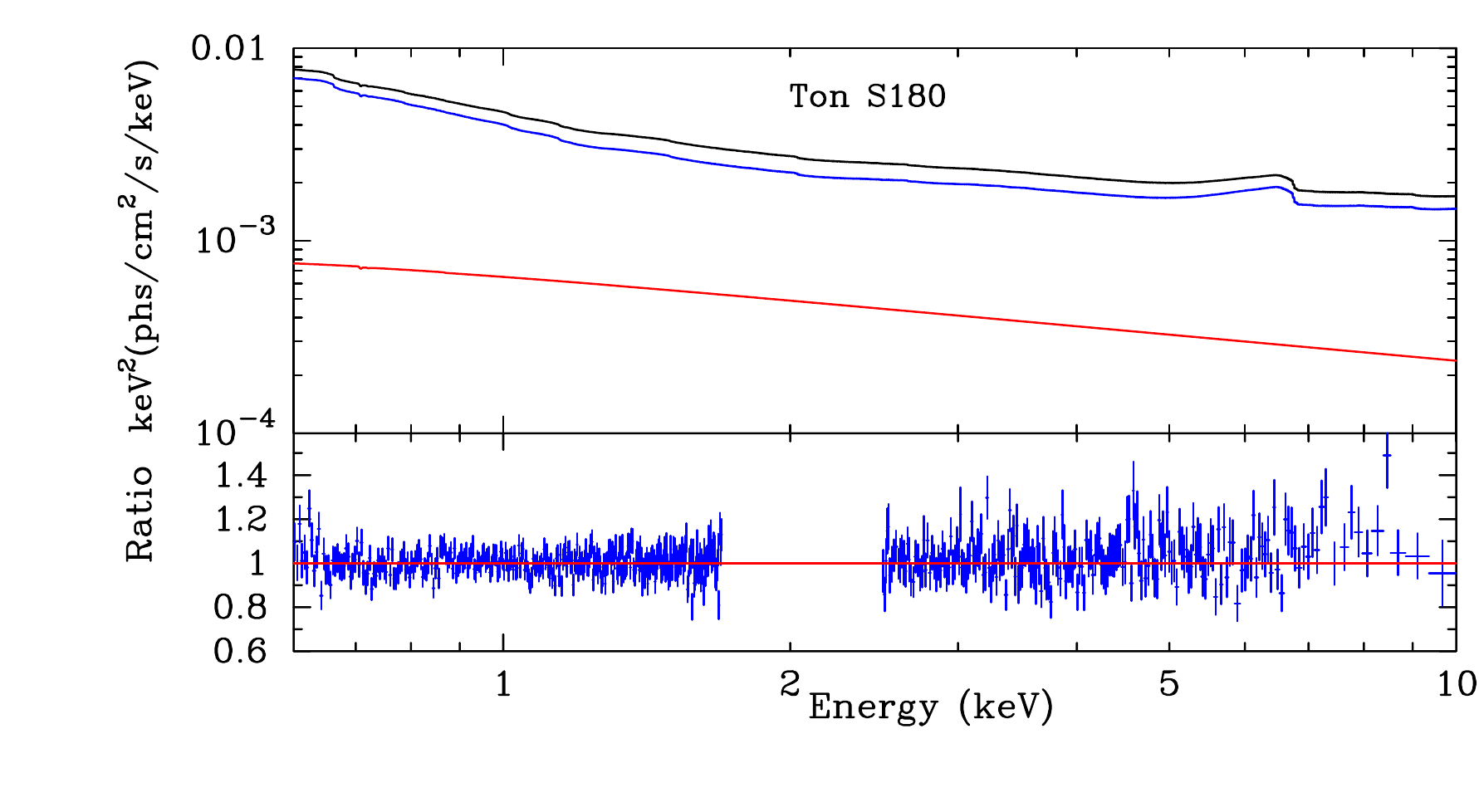}
\includegraphics[width=8.5cm,trim={0.5cm 0 3cm 18cm},clip]{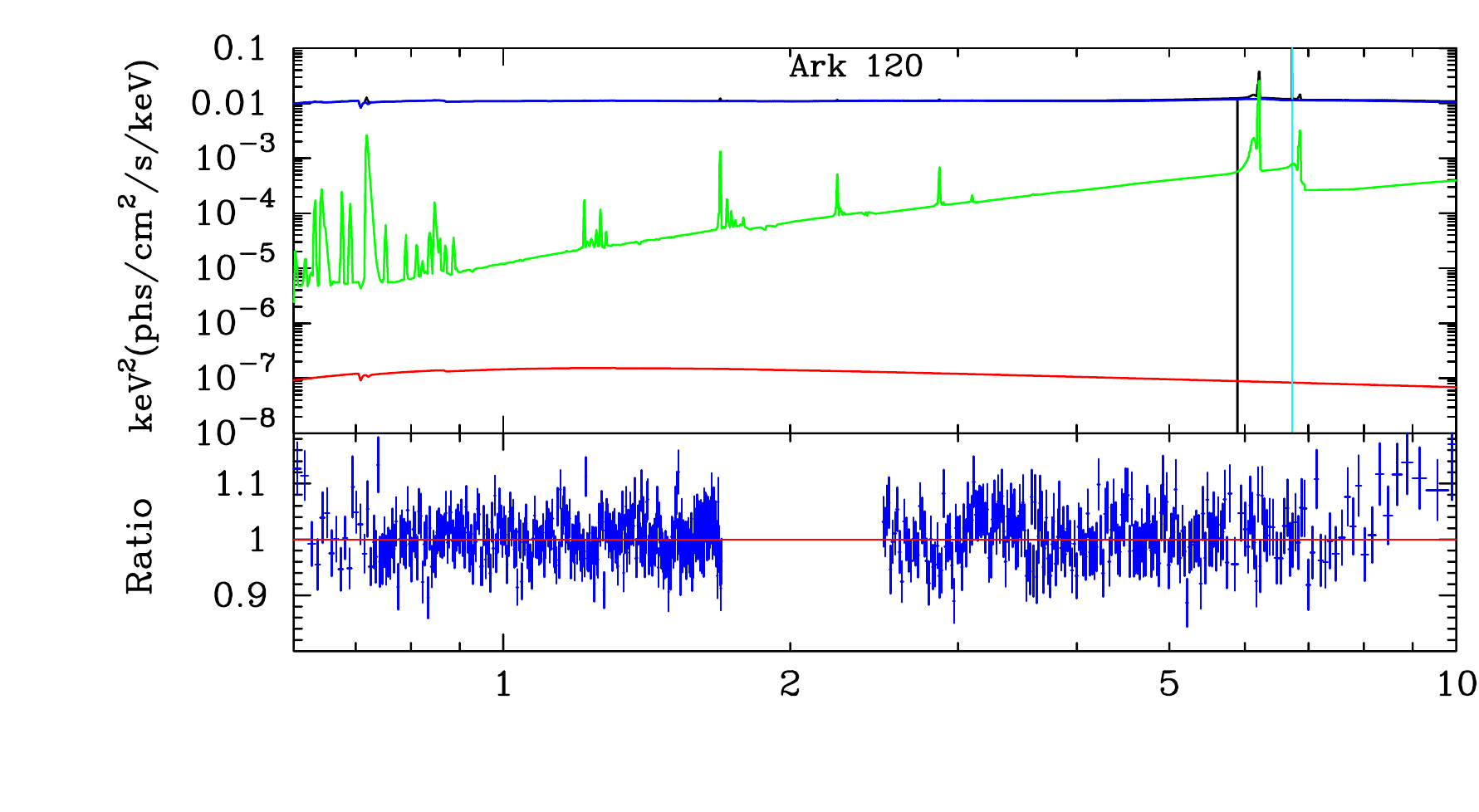} \\
\includegraphics[width=8.5cm,trim={0.5cm 0 3cm 18cm},clip]{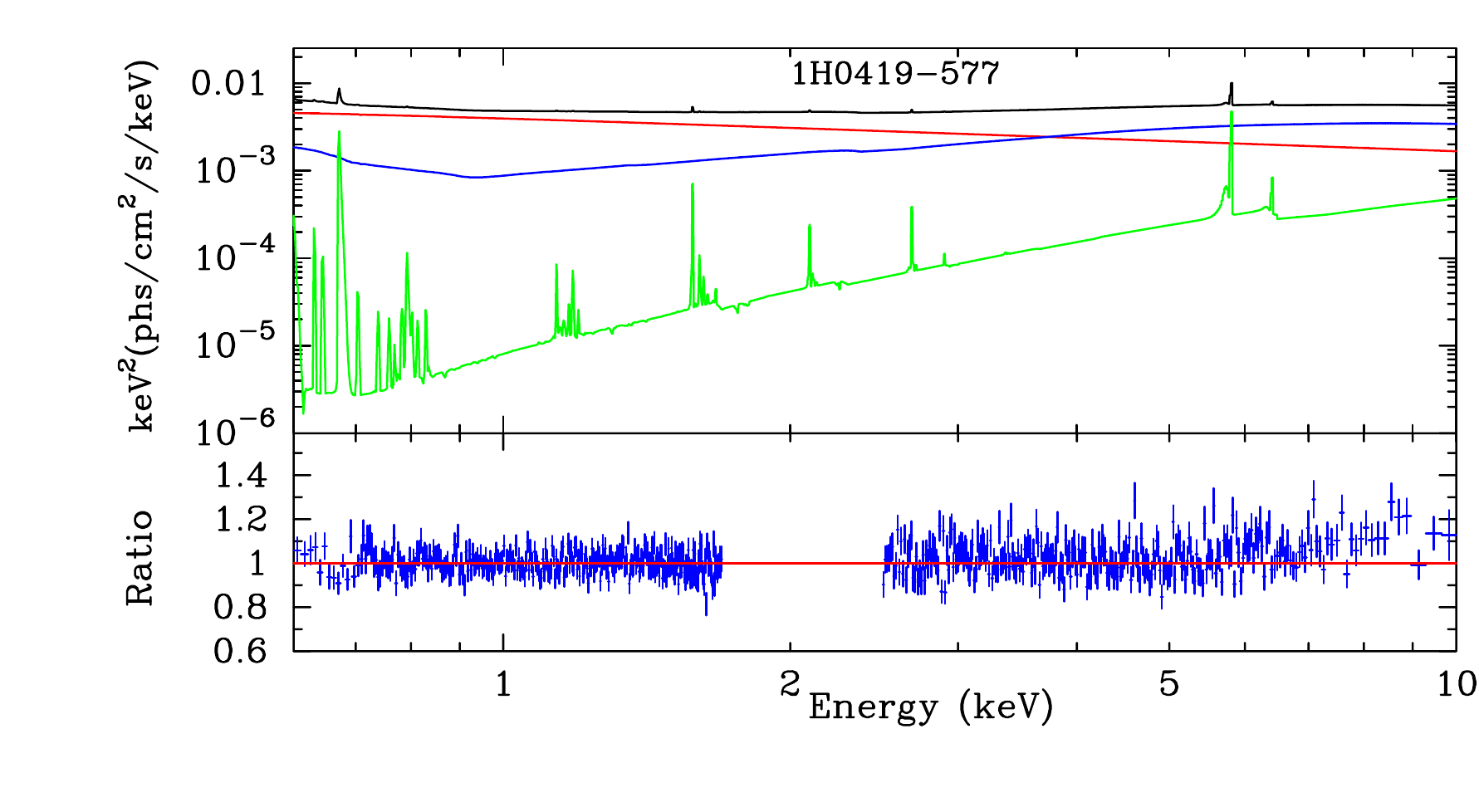}
\includegraphics[width=8.5cm,trim={0.5cm 0 3cm 18cm},clip]{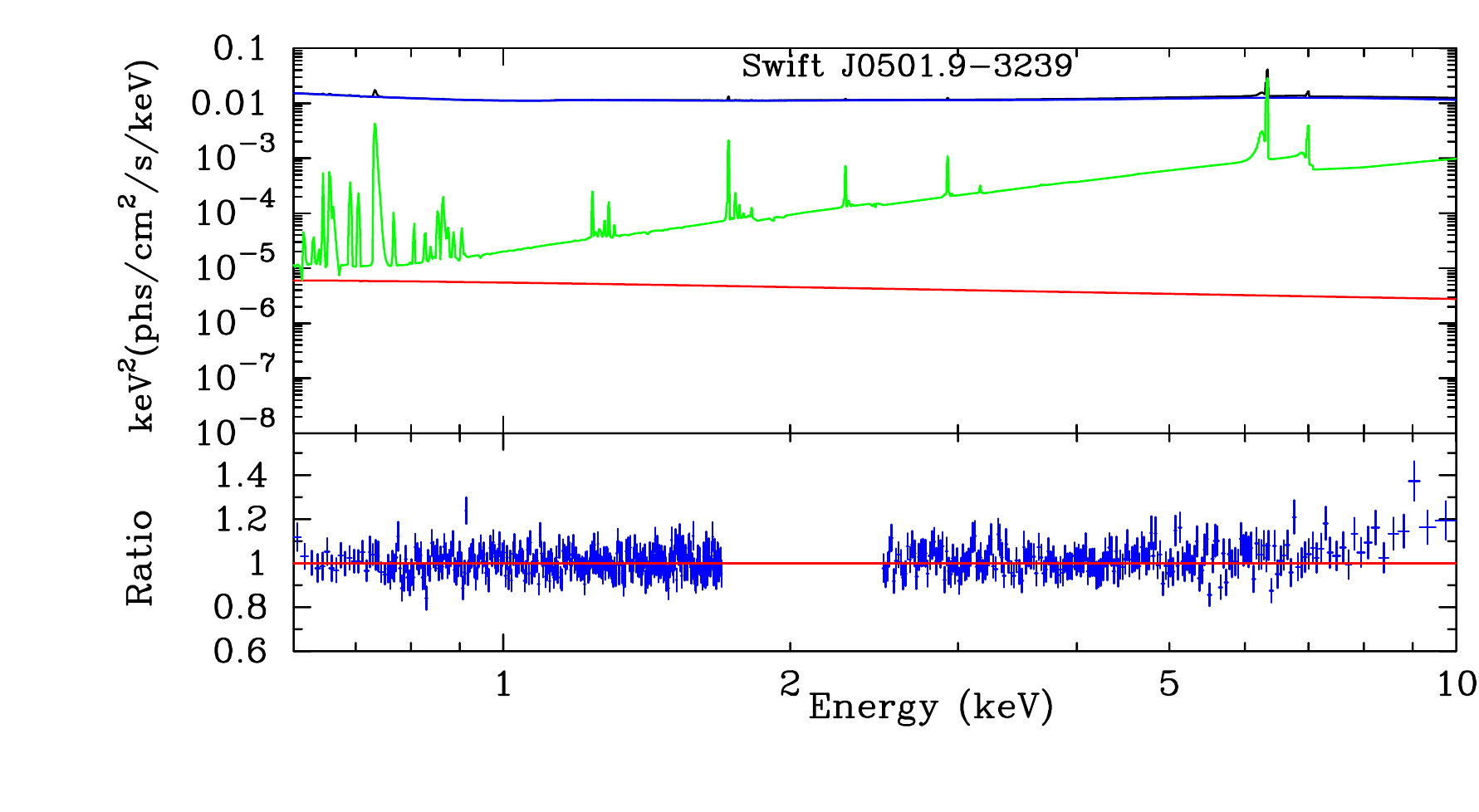}
\end{center}
\vspace{-0.7cm}
\caption{Spectra of the best fit models with the corresponding components (upper panels) and data to best-fit model ratios (lower panels) for the four sources of our study. The total spectra are in black, the power law components from the coronas are in red, the relativistic reflection components from the disks are in blue, the non-relativistic reflection components from distant reflectors are in green, the narrow lines are in cyan. Note that in the case of Ark~120 and Swift~J0501.9--3239 the total flux (black) and the relativistic reflection component (blue) almost overlap. \label{f-r}}
\end{figure*}

\begin{figure*}[t]
\begin{center}
\includegraphics[width=8.5cm,trim={0.5cm 1.2cm 0.5cm 2cm},clip]{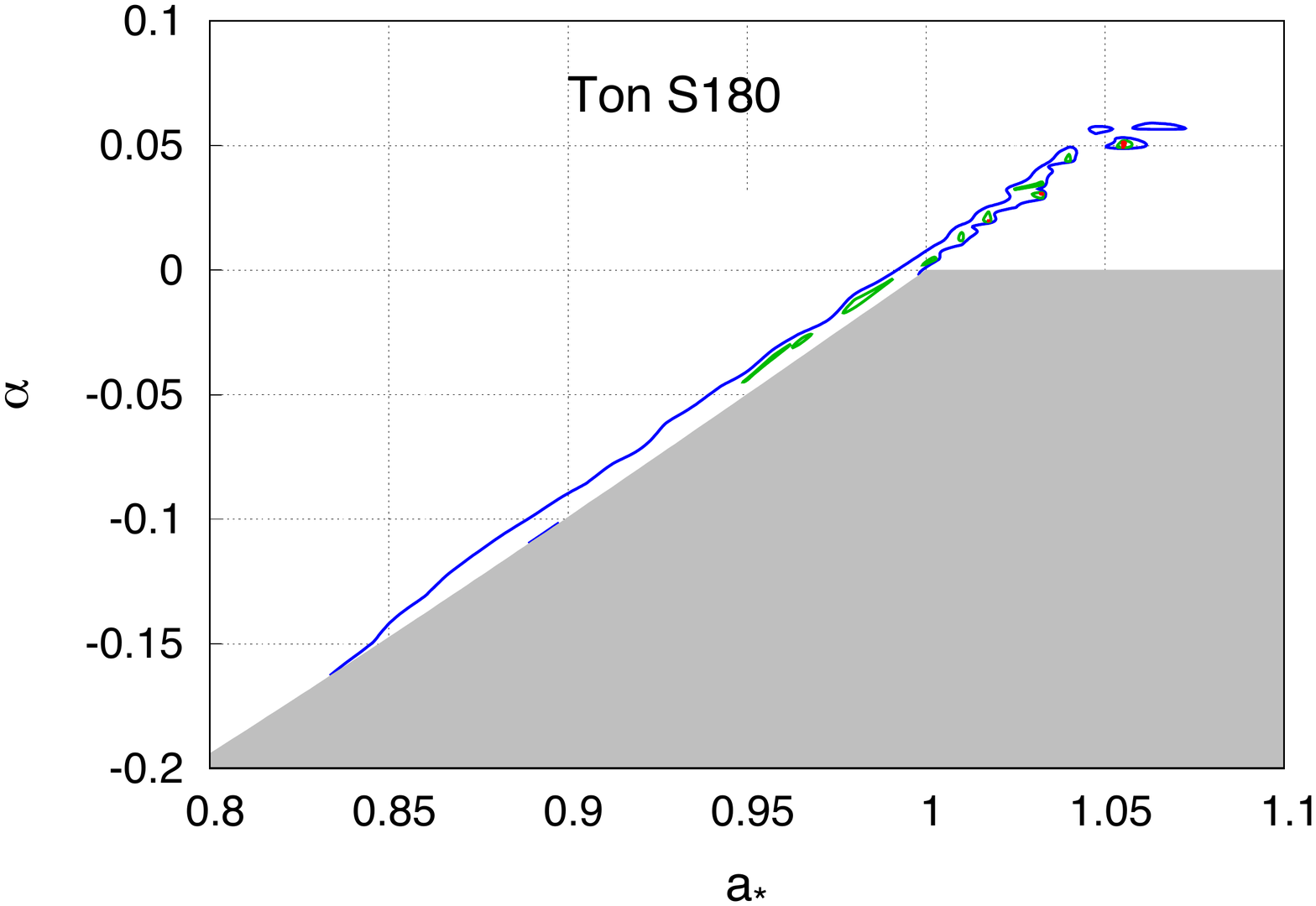}
\includegraphics[width=8.5cm,trim={0.5cm 1.2cm 0.5cm 2cm},clip]{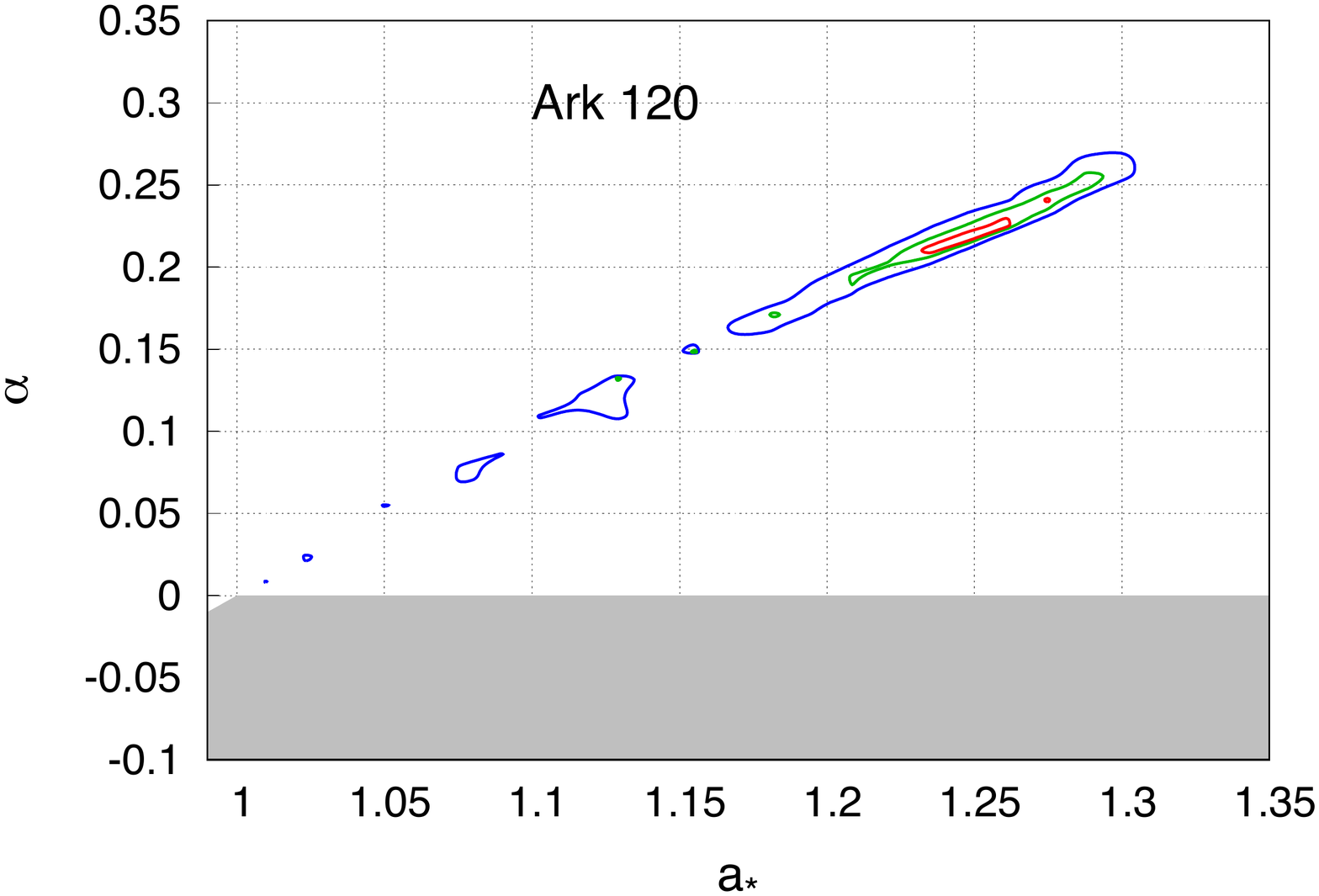} \\
\includegraphics[width=8.5cm,trim={0.5cm 1.2cm 0.5cm 2cm},clip]{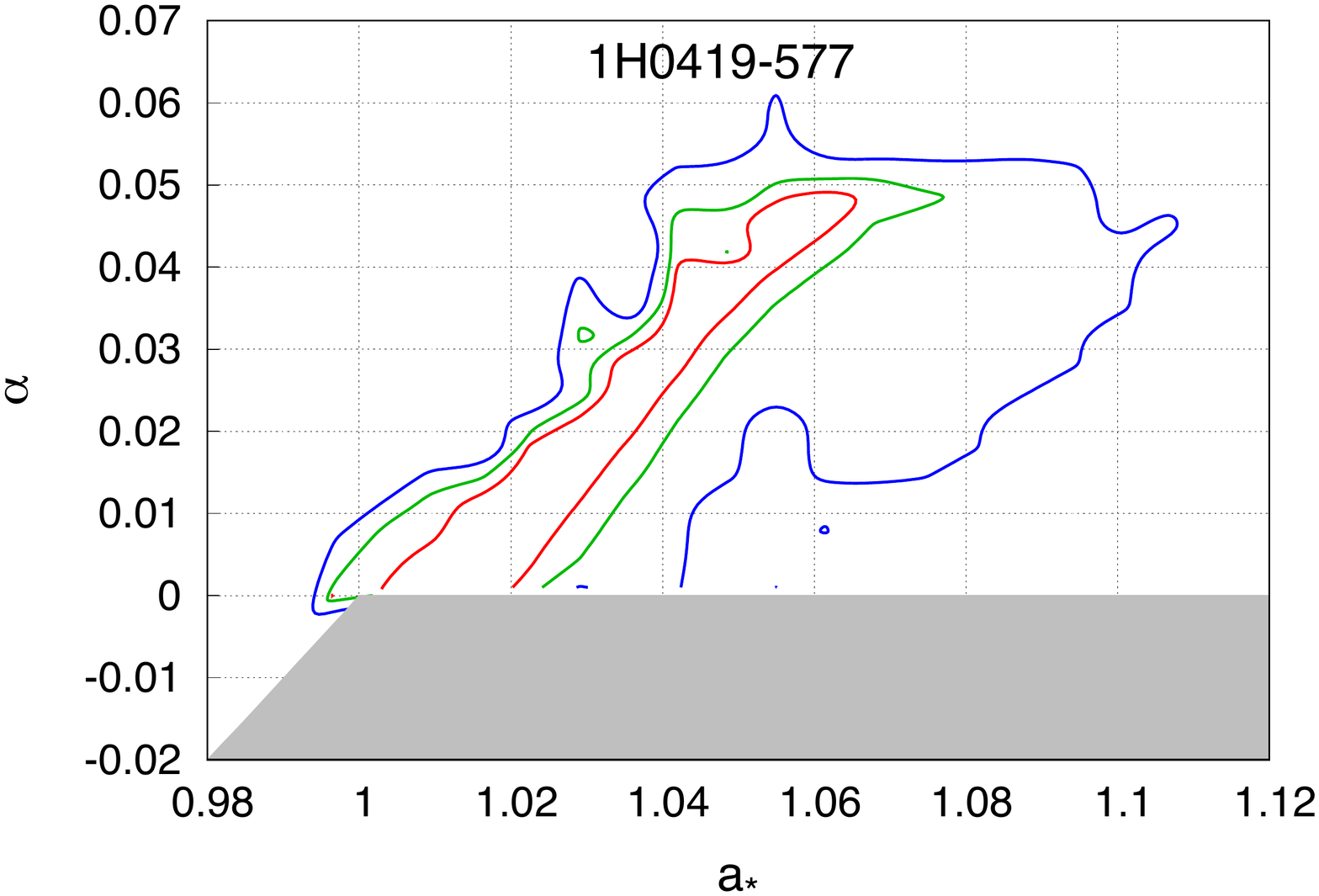}
\includegraphics[width=8.5cm,trim={0.5cm 1.2cm 0.5cm 2cm},clip]{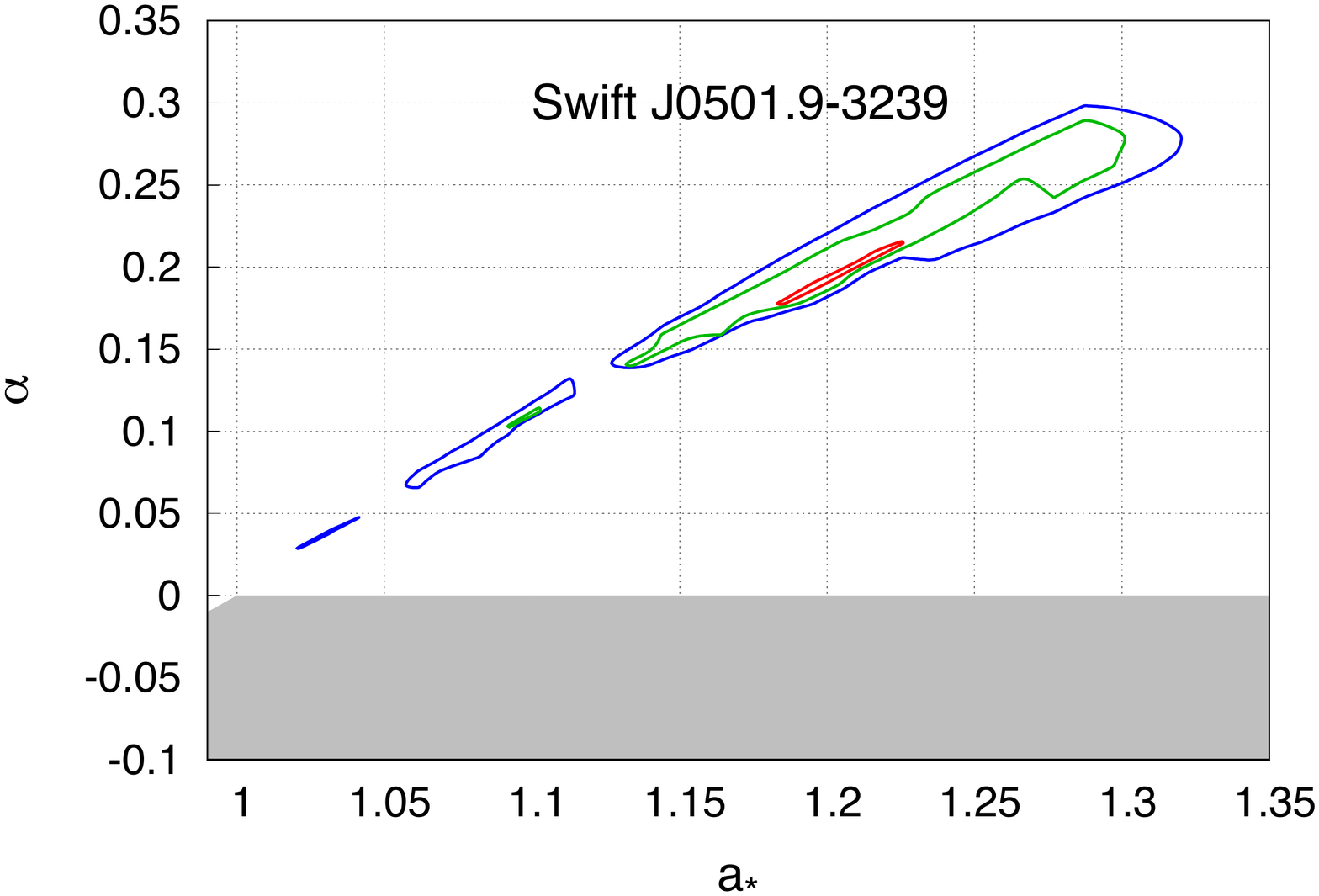}
\end{center}
\vspace{-0.9cm}
\caption{Constraints on the spin parameter $a_*$ and the deformation parameter $\alpha$ for the four sources of our study. The red, green, and blue curves correspond, respectively, to the 68\%, 90\%, and 99\% confidence level limits for two relevant parameters ($\Delta\chi^2 = 2.30$, 4.61, and 9.21, respectively). The gray region is ignored in our study because it is the parameter space with naked singularities. \label{f-c}}
\end{figure*}

The results of our fits are reported in Tab.~\ref{t-fit}. Every source has two columns. The first column shows the fit of the Kerr model ($a_* \le 0.998$ and $\alpha = 0$). The second column is for the fit with $a_*$ and $\alpha$ free. All the uncertainties are at the 90\% confidence level for one relevant parameter ($\Delta\chi^2 = 2.71$). $^*$ indicates that the parameter is frozen in the fit. The spectra of the best-fit model with the corresponding components and the data to best-fit model ratios for the case of free $a_*$ and $\alpha$ are reported in Fig.~\ref{f-r}.

Fig.~\ref{f-c} shows the constraints on the spin parameter vs deformation parameter plane for the four sources when both $a_*$ and $\alpha$ are free in the fit. The red, green, and blue curves correspond, respectively, to the 68\%, 90\%, and 99\% confidence level limits for two relevant parameters (i.e. $\Delta\chi^2 = 2.30$, 4.61, and 9.21, respectively). In the gray region, the spacetimes have no black holes but naked singularities, and are ignored in our analysis. Note that these constraints on $a_*$ and $\alpha$ are obtained after marginalizing over all the free parameters in the fit. Since there are several free parameters, the algorithm of XSPEC has some problems to scan the parameter space and find the right minimum of $\chi^2$ at every point of the plane $a_*$ vs $\alpha$. This causes some islands in the contour plots in Fig.~\ref{f-c}. If we could better scan the parameter space, the 68\%, 90\%, and 99\% confidence level regions in Fig.~\ref{f-c} should slightly increase in size and these islands should disappear.

\begin{figure*}[t]
\begin{center}
\includegraphics[width=8.5cm]{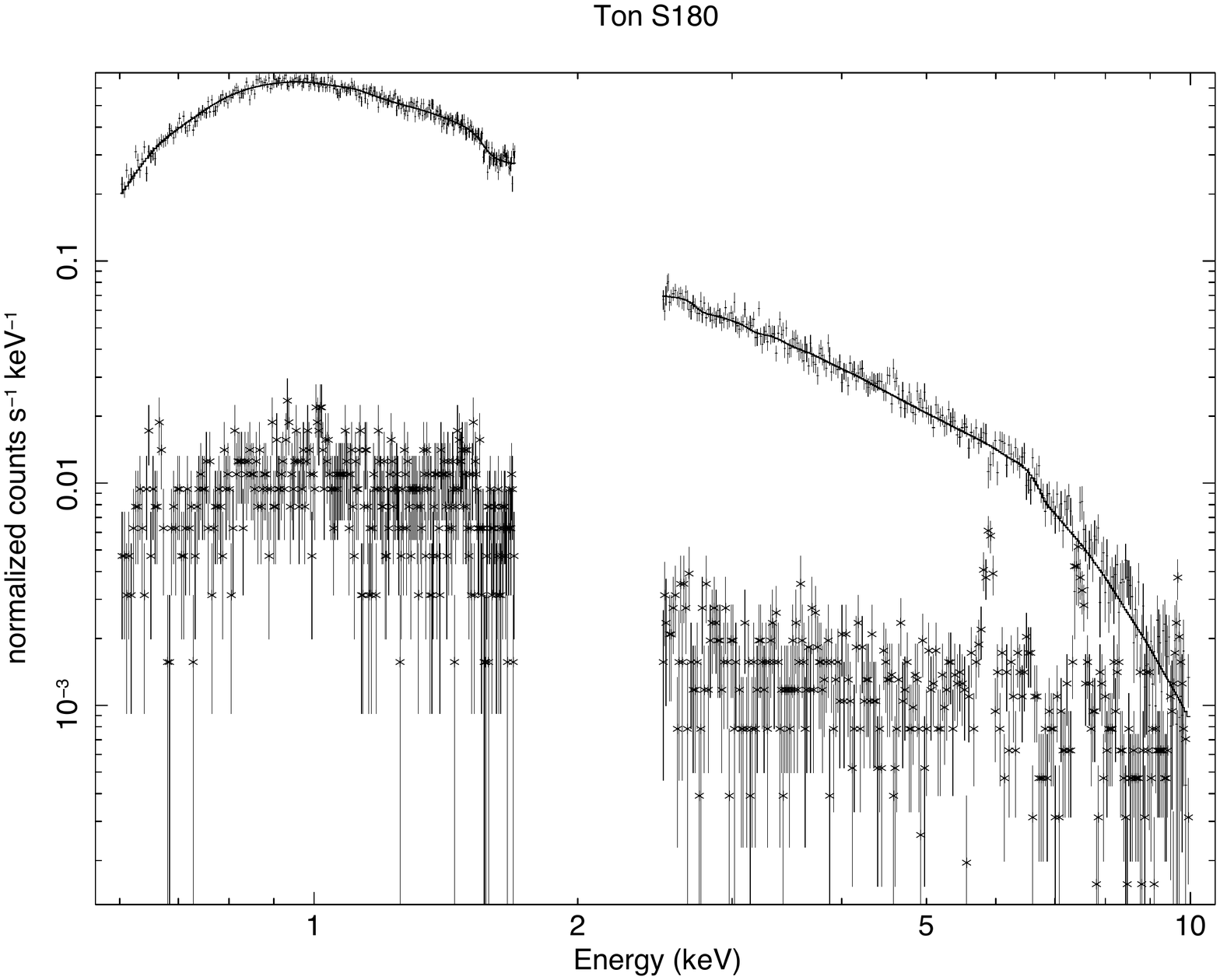}
\includegraphics[width=8.5cm]{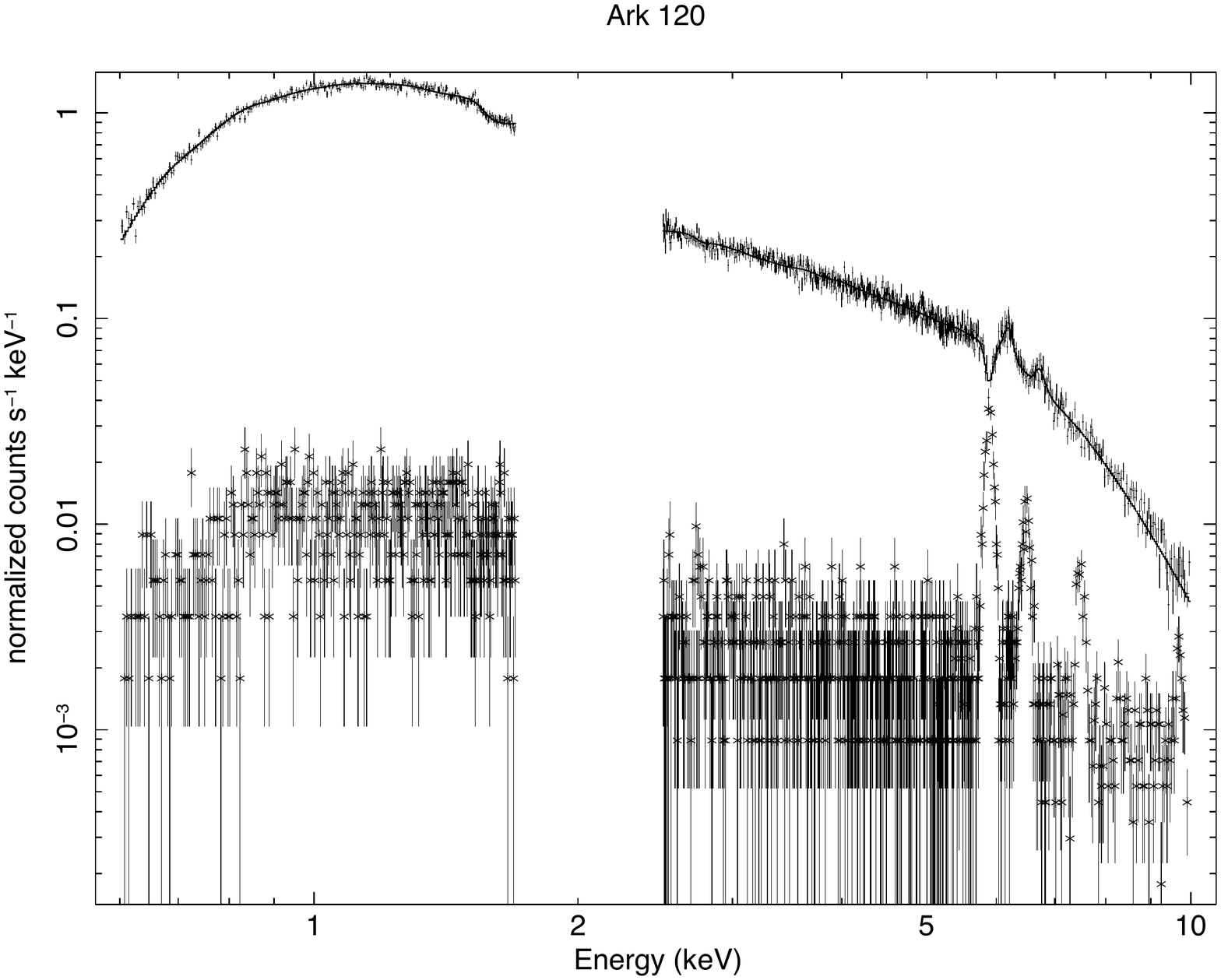} \\
\vspace{-0.4cm}
\includegraphics[width=8.5cm]{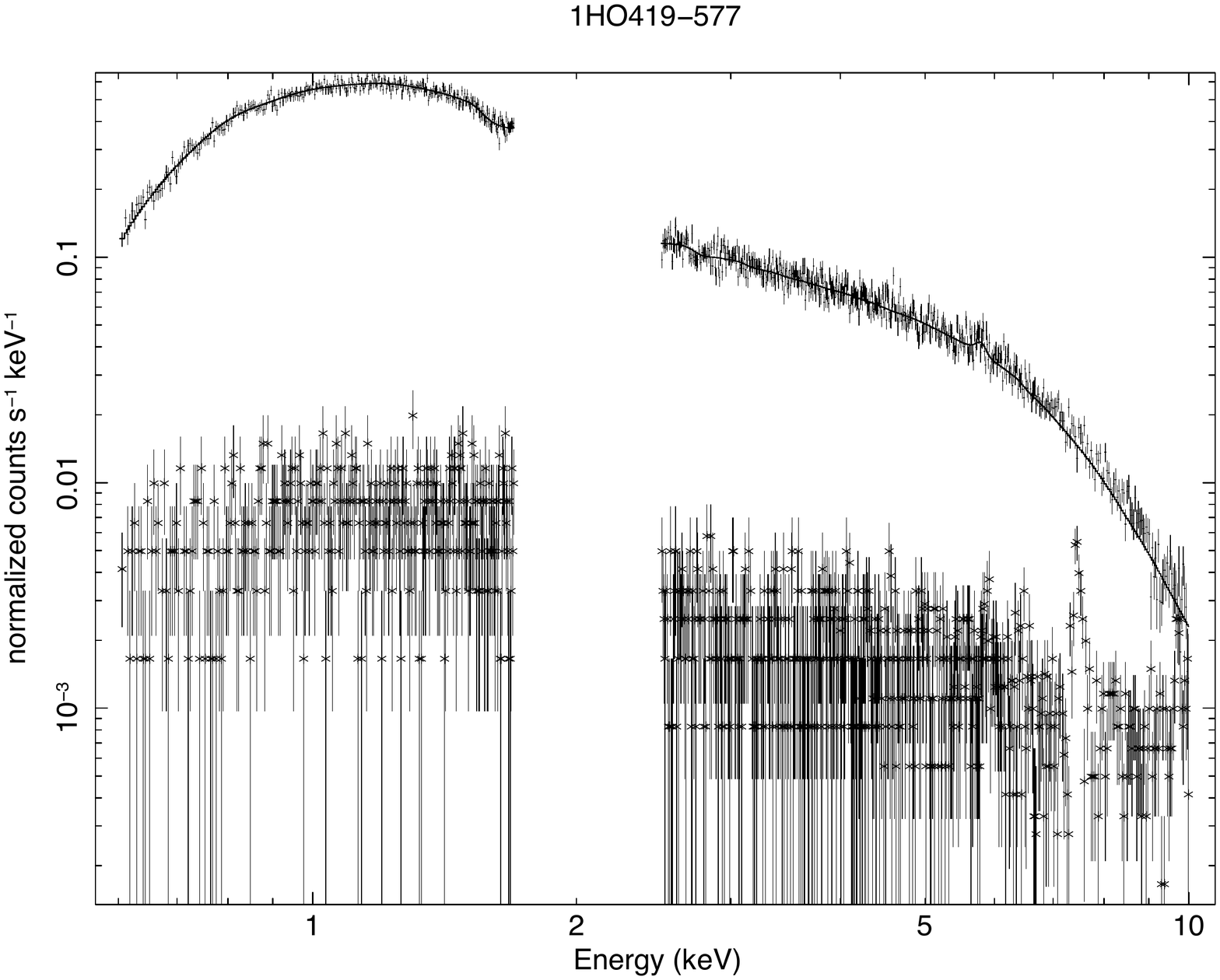}
\includegraphics[width=8.5cm]{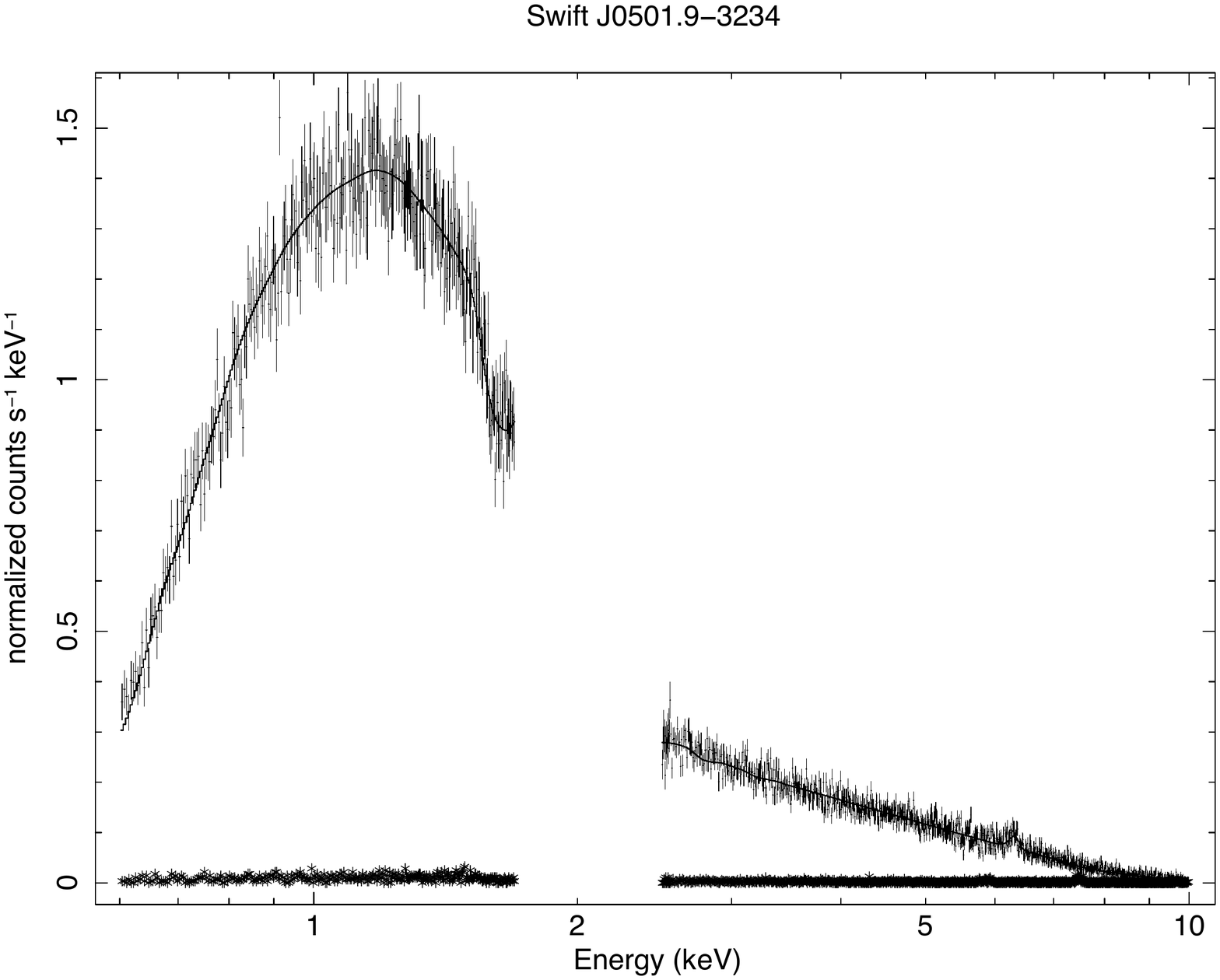}
\end{center}
\vspace{-0.8cm}
\caption{{Source spectra, background spectra, and folded models of the \textsl{Suzaku} data of Ton~S180, Ark~120, 1H0419--577, and Swift~J0501.9--3239.} \label{f-bg}}
\end{figure*}

\begin{figure*}[t]
\begin{center}
\includegraphics[width=8.5cm,trim={0.5cm 0.0cm 2.5cm 12.5cm},clip]{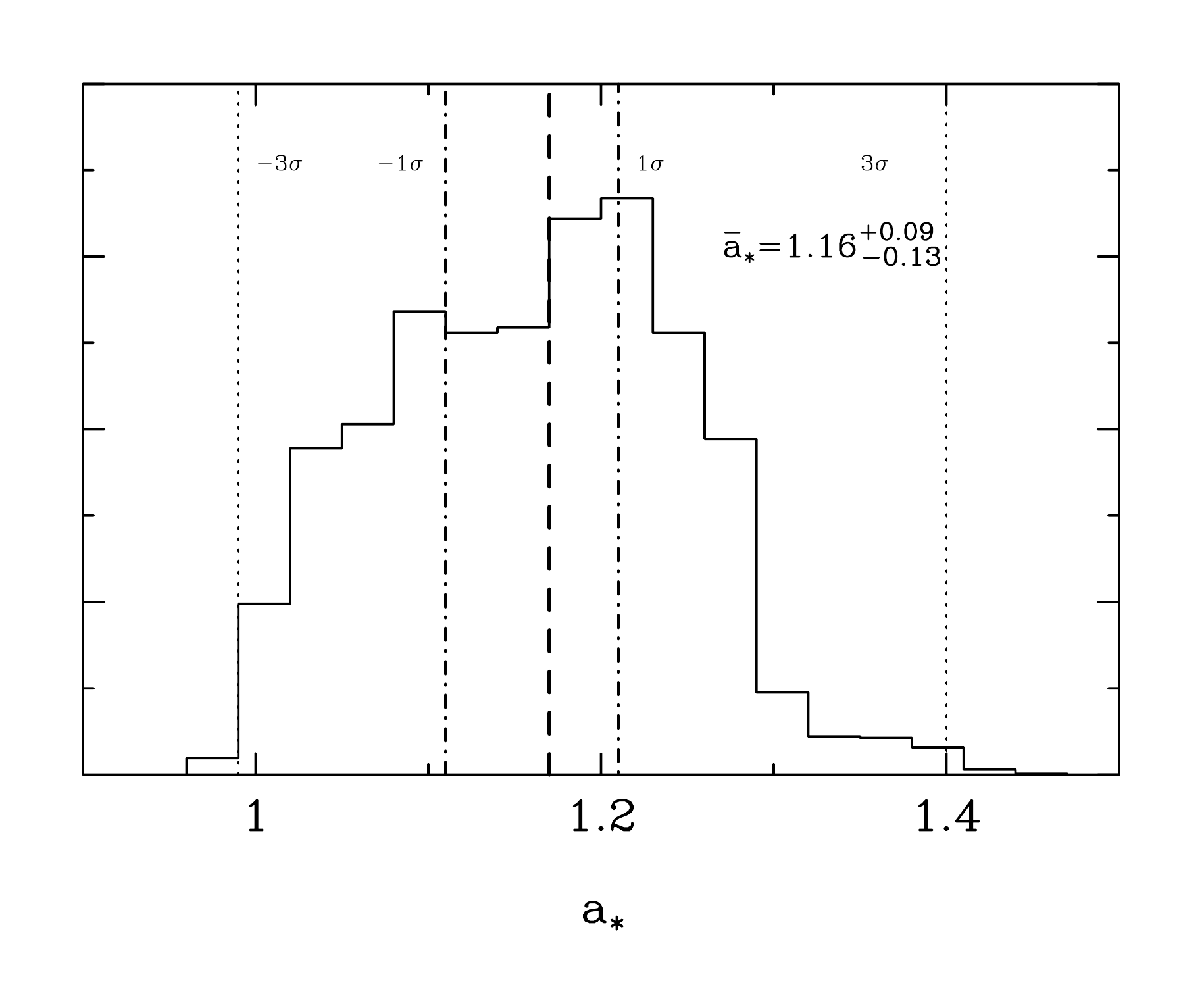}
\includegraphics[width=8.5cm,trim={0.5cm 0.0cm 2.5cm 12.5cm},clip]{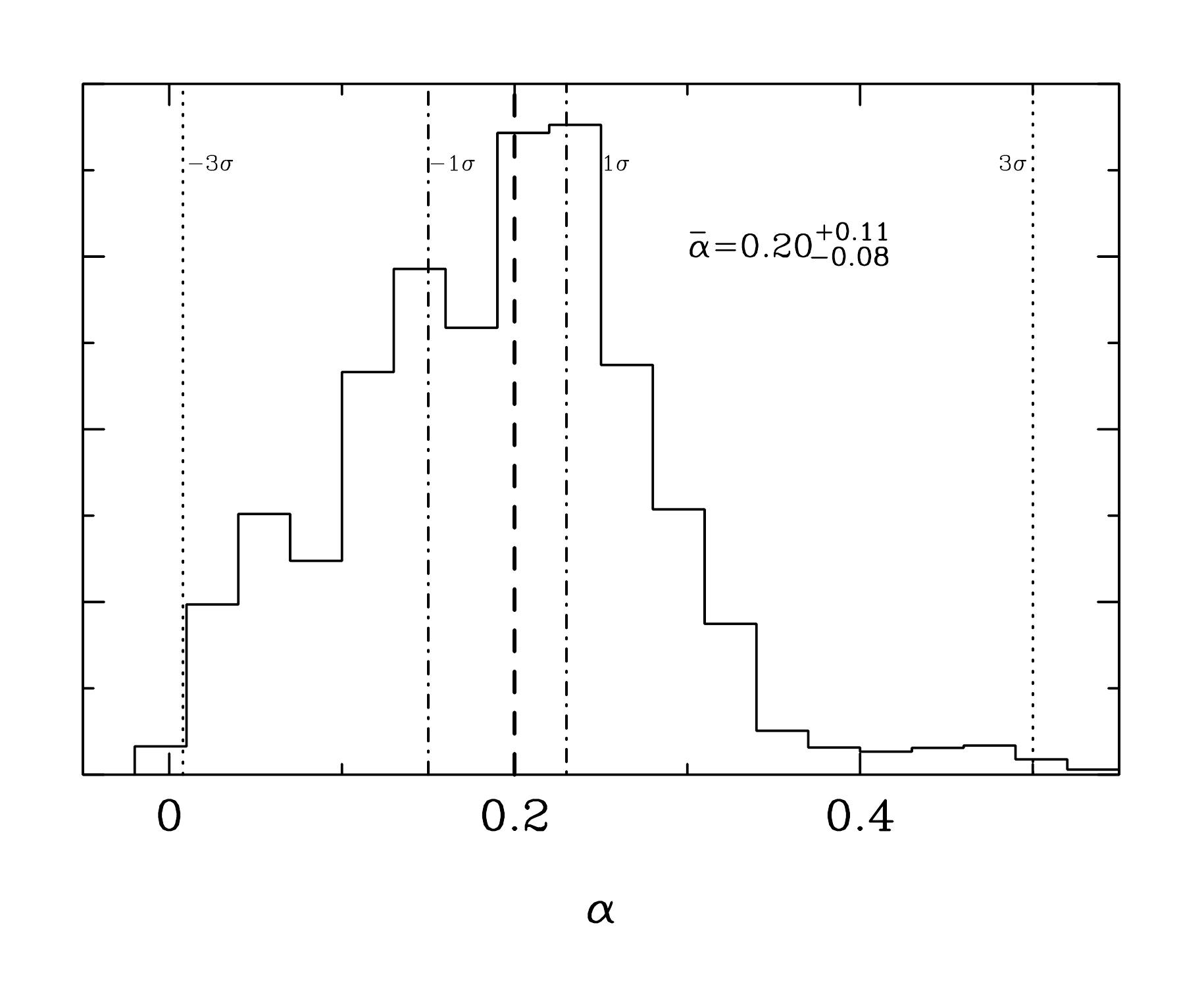}
\end{center}
\vspace{-0.9cm}
\caption{{Histograms of the spin parameter $a_*$ (left panel) and of the deformation parameter $\alpha$ (right panel) for Ark~120 from our MCMC analysis. The thick-dashed vertical lines mark the mean value of the fitting parameter, while the thin-dotted-dashed and thin-dotted vertical lines mark, respectively, the 1-$\sigma$ and 3-$\sigma$ limits.} \label{f-mcmc1}}
\vspace{0.3cm}
\begin{center}
\includegraphics[width=8.5cm,trim={0.5cm 0.0cm 2.5cm 12.5cm},clip]{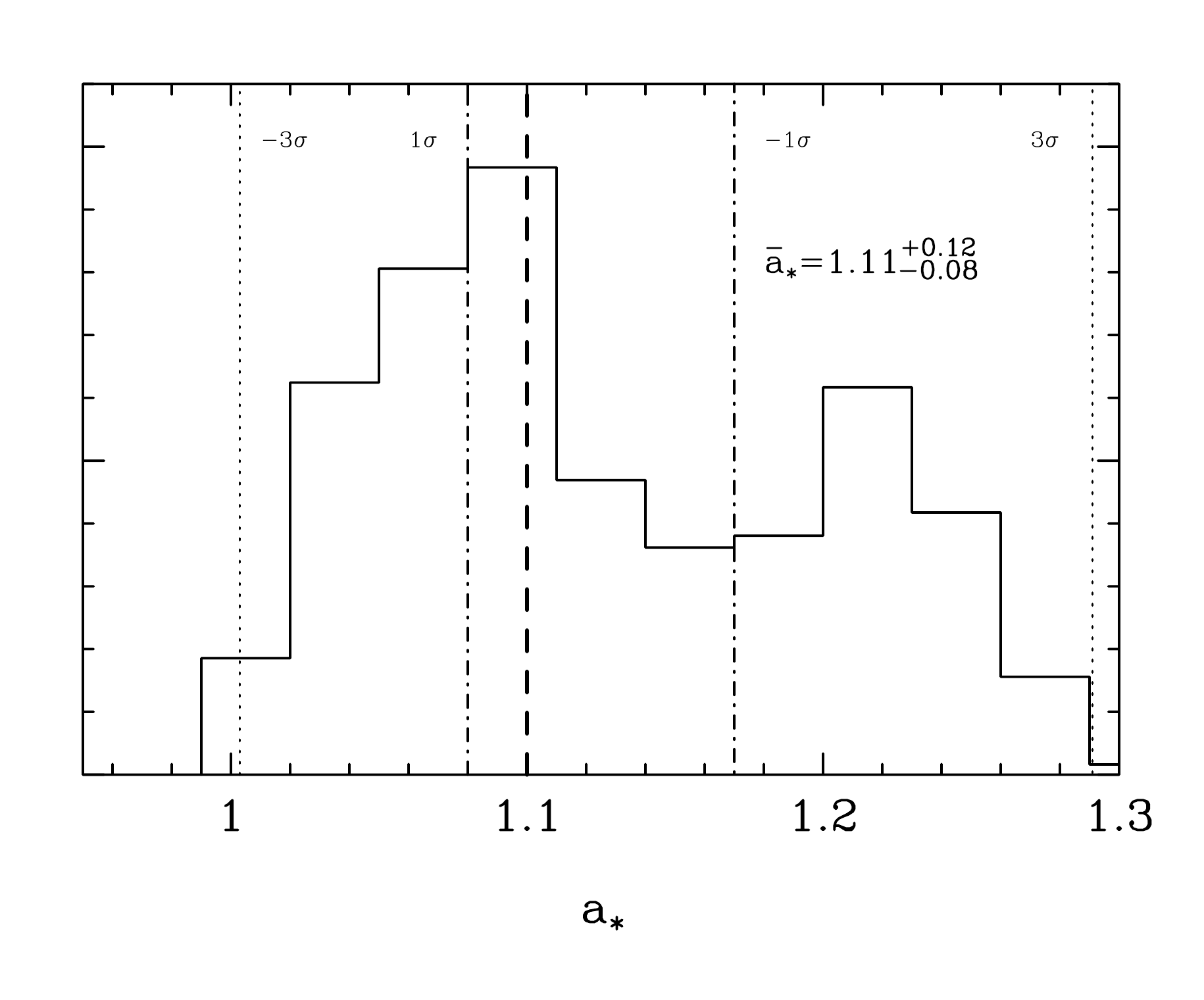}
\includegraphics[width=8.5cm,trim={0.5cm 0.0cm 2.5cm 12.5cm},clip]{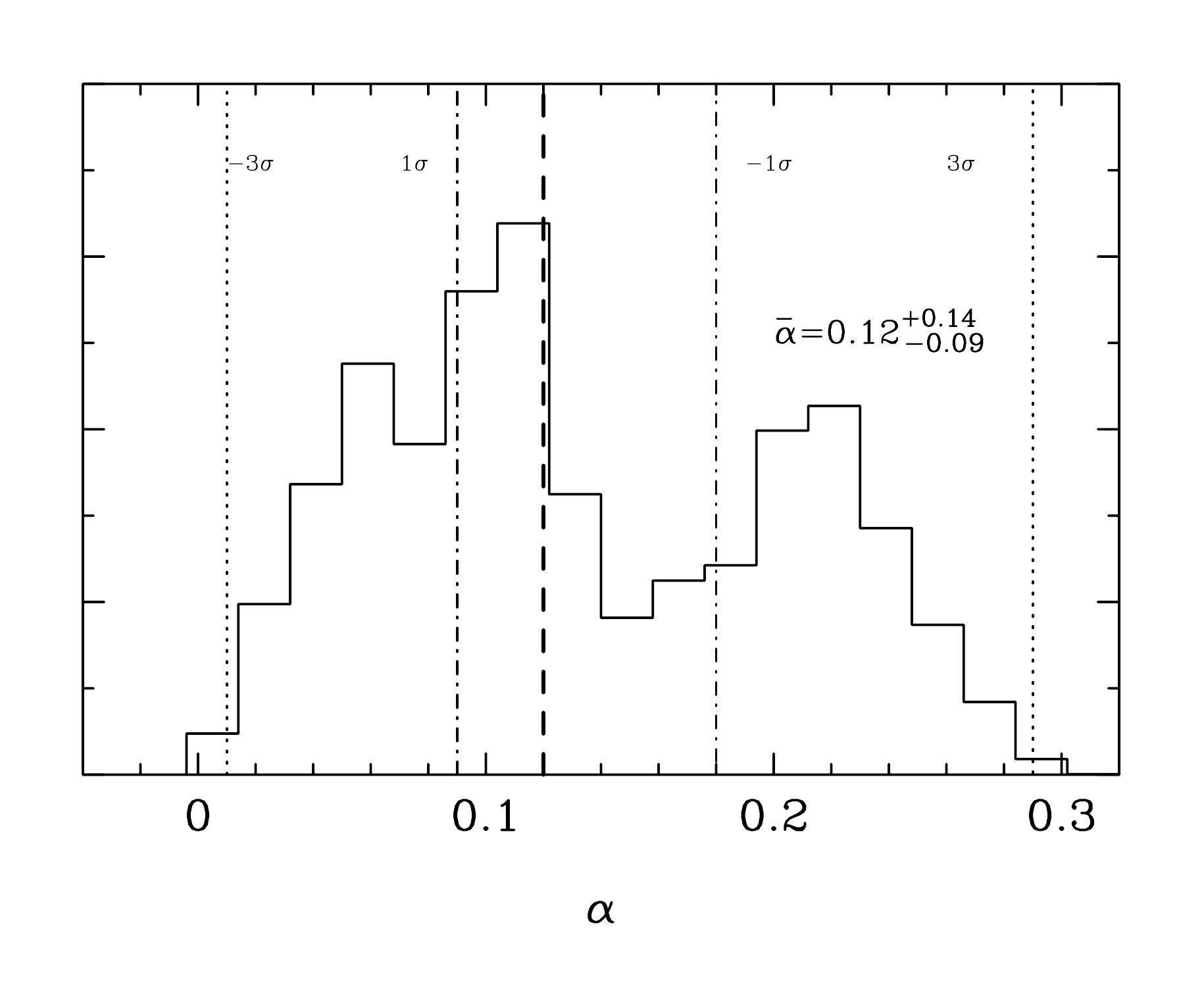}
\end{center}
\vspace{-0.9cm}
\caption{{Histograms of the spin parameter $a_*$ (left panel) and of the deformation parameter $\alpha$ (right panel) for Swift~J0501.9--3239 from our MCMC analysis. The thick-dashed vertical lines mark the mean value of the fitting parameter, while the thin-dotted-dashed and thin-dotted vertical lines mark, respectively, the 1-$\sigma$ and 3-$\sigma$ limits.} \label{f-mcmc2}}
\end{figure*}

%%%%%%%%%%%%%%%%%%%%%%%%%%%%%%%

\section{Discussion and conclusions \label{s-dc}}

In Ref.~\cite{bare}, we tested the Kerr hypothesis with seven supermassive black holes by analyzing some \textsl{Suzaku} observations with {\sc relxill\_nk} and constraining the deformation parameters $\alpha_{13}$ and $\alpha_{22}$ of the Johannsen metric. For all sources, our study found that the spacetime metric around the central compact object was consistent with the Kerr solution, with remarkably strong constraints on the deformation parameters $\alpha_{13}$ and $\alpha_{22}$. For several sources, the best-fit value of the spin parameter was stuck at the maximum value of the model, which is 0.998, indicating that the inner edge of the accretion disk of these sources is very close to the compact object, and photons from that region are strongly redshifted. Since such an effect usually gets weaker when we deform the Kerr metric, it is quite automatic that similar sources can provide stringent constraints on the values of most deformation parameters. However, best-fit values stuck at the boundary of the parameter space of a model may also look suspicious, indicating a possible breakdown of the model itself.

Motivated by such a doubt, in the present work we have reanalyzed those suspicious sources with a different parametric black hole spacetime. In particular, we chose a metric in which black holes can have a spin parameter $a_*$ exceeding 1 and the radiation from the ISCO radius can be more redshifted than in the Kerr background. Not surprisingly, the best-fit values that were stuck at 0.998 in Ref.~\cite{bare} moved to $a_* > 1$ for all sources. Our results are summarized in Tab.~\ref{t-fit}, where the last row shows the difference between the minimum of $\chi^2$ between the Kerr black hole model and the model with free deformation parameter.

For Ton~S180 and 1H0419--577, the new analysis is still consistent with the hypothesis that the spacetime metric around the two compact objects is described by the Kerr metric at 90\% confidence level (i.e. $\Delta\chi^2 < 2.71$ between the Kerr and non-Kerr model). In other words, we have a new test, with a different deformation parameter, confirming the Kerr nature of these objects. For Swift~J0501.9--3239, the comparison between the Kerr and non-Kerr model gives $\Delta\chi^2 = 6.86$. For Ark~120, the Kerr solution is only recovered if we consider a higher confidence level ($\Delta\chi^2 = 20.40$), indicating that the non-Kerr model provides a better fit.

From Fig.~\ref{f-r}, we can notice that the best-fit models have positive residuals above 8~keV, in particular for Ark~120 and Swift~J0501.9--3239. This is not due to the background, as we can see from Fig.~\ref{f-bg}, but it is difficult to determine its exact origin. It might be due to modeling uncertainties. For example, positive residuals above 8~keV may be caused by partial Comptonisation of the reflection spectrum by an extended corona above the accretion disk~\cite{Wilkins:2014caa}. If so, as well as for similar cases, we may expect that such modeling uncertainties do not significantly affect the measurements of the spin and the deformation parameters, which are mainly sensitive to the iron line region and the soft excess. The existence of high quality high energy data would surely help to fit the reflection component better, but, as pointed out in the previous section, the available data from the PIN instrument are of poor quality.

A comparison with published analyses of the same sources is not straightforward because they are obtained from different observations and/or employing different models. The simplest comparison is with the study reported in~\cite{w13}, where the authors analyze the same \textsl{Suzaku} observations but with a different reflection model ({\sc reflionx}~\cite{Ross:2005dm}, which is transformed into a relativistic spectrum with {\sc relconv}~\cite{Dauser:2010ne}). The convolution of a non-relativistic reflection model with {\sc relconv} does not properly take into account the emission angle at every point of the disk, and this leads to underestimating the black hole spin parameter when we assume the Kerr metric~\cite{inp}. This is indeed our case, as the authors of Ref.~\cite{w13} do not find extremely high spins for Ton~S180 and Ark~120.

More detailed studied of these sources are reported in other publications from the analysis of different data. For example, an analysis of Swift~J0501.9--3239 is reported in Ref.~\cite{Agis-Gonzalez:2014nja}, where the authors conduct a multi-epoch spectral analysis including also the 2008 \textsl{Suzaku} observation considered in our work. They employ a different relativistic reflection model and include different data, so a discrepancy in the value of the inclination angle between our best fit ($i < 7$~deg) and their result ($i \sim 50$~deg) can likely be related to modeling uncertainties that would act in a different way for different observations even if we used the same models.

Considering the difficulties of the $\chi^2$ minimizing algorithm of XSPEC to reliably find a minimum and the uncertainties in complicated $\chi^2$ landscape, we performed a Markov-Chain-Monte-Carlo (MCMC) analysis for Ark~120 and Swift~J0501.9--3239, for which the non-Kerr solution provides a significantly better fit. We followed the MCMC analysis of Ref.~\cite{Kara:2017jdb}. We used the {\sc xspec\_emcee} code by Jeremy Sanders\footnote{available on github, https://github.com/jeremysanders/xspec\_emcee .} which uses {\sc emcee} (MCMC Ensemble sampler implementing Goodman \& Weare algorithm) to analyze spectral data within XSPEC. Here, we used 10,000 iterations with 50 walkers and burnt the first 1,000 steps. For Ark~120, our MCMC analysis finds (90\% confidence level)
\be
a_* = 1.16_{-0.13}^{+0.09} \, , \quad \alpha = 0.20_{-0.08}^{+0.11} \, .
\ee
For Swift~J0501.9--3239, our measurements of $a_*$ and $\alpha$ is (90\% confidence level)
\be
a_* = 1.11_{-0.08}^{+0.12} \, , \quad \alpha = 0.12_{-0.14}^{+0.09} \, .
\ee
The histograms are shown in Fig.~\ref{f-mcmc1} and Fig.~\ref{f-mcmc2}, respectively for Ark~120 and Swift~J0501.9--3239. The MCMC analysis thus confirms that the non-Kerr solution is preferred for both sources. The Kerr metric is not recovered at 3-$\sigma$.

While it is definitively suggestive the possibility of finding new physics, we have good reasons to suspect that systematic uncertainties (broadly defined, not taken into account in our analysis) exceed the statistical uncertainties, thus leading to a wrong measurement of the deformation parameter.
The modeling simplifications in {\sc relxill\_nk} inevitably lead to a number of systematic uncertainties currently not under control, see Refs.~\cite{hh,r3} for their list and possible impact. As stressed in~\cite{hh,r3}, it is extremely important to select the right sources for limiting all these systematic uncertainties. For example, {\sc relxill\_nk} employs the Novikov-Thorne model for geometrically thin and optically thick disks and assumes that the inner edge is at the ISCO radius. However, we know that geometrically thin disks with inner edge at the ISCO require that the source is in the soft state with an accretion luminosity between a few percent to about 30\% of its Eddington limit, while most supermassive black holes have much higher accretion rates, see Tab.~1 in \cite{brenneman}. Ton~S180 and 1H0419--577 seem to accrete near the Eddington limit, which should induce modeling bias in the estimate of some parameters~\cite{Riaz:2019bkv,Riaz:2019kat}. For Swift~J0501.9--3239, there is no estimate of its accretion luminosity in Eddington units. The estimated accretion luminosity of Ark~120 is at the level of a few percent, and thus consistent with the thin disk model, so a too high accretion luminosity should not be the explanation of the measured non-vanishing $\alpha$ for this source.

Among the possible systematic uncertainties affecting the measurement of accreting black holes using X-ray reflection spectroscopy, our previous studies showed that the choice of the model of the intensity profile can be crucial in the final measurement of the deformation parameters of the spacetime~\cite{hh,yuexin,yuexin2}. It is possible that a simple power-law or a broken power-law are not enough to fit the \textsl{Suzaku} data of Ark~120, and this may cause the apparent detection of a non-vanishing $\alpha$. The model has many other simplifications (e.g., constant ionization parameter and electron density over the whole disk, no emission from the plunging region between the inner edge of the disk and the black hole, disk with Solar metallicity except for iron, reflection spectrum calculated employing non-relativistic formulas for Compton scattering, etc.), but their impact on the estimate of the model parameters is usually smaller than the effect produced by the choice of the emissivity profile.

Lastly, we also not that the quality of the \textsl{Suzaku} data analyzed in the present work is not excellent, and this limits a more detailed study. In particular, we do not have high energy data. It is not surprising that our current best tests of the Kerr metric are obtained from \textsl{Suzaku} data of GRS~1915+105~\cite{yuexin2}, where the brightness of the source permits to use the PIN data as well, and from simultaneous observations of \textsl{XMM-Newton} and \textsl{NuSTAR} of MCG--6--30--15~\cite{noi}, where we have both a good energy resolution near the iron line with \textsl{XMM-Newton} and a broad energy band with \textsl{NuSTAR}.

In conclusion, our work confirms that it is extremely important to select well-understood sources for testing the Kerr hypothesis and, in general, to get reliable measurements of the parameter of a system. While apparent near-extremal Kerr black holes can provides quite stringent constraints on a number of deformation parameters, it is always dangerous to rely on sources in which some best-fit values are stuck at the boundary of a model.

%%%%%%%%%%%%%%%%%%%%%%%%%%%%%%%

\vspace{0.5cm}

{\bf Acknowledgments --}
This work was supported by the Innovation Program of the Shanghai Municipal Education Commission, Grant No.~2019-01-07-00-07-E00035, the National Natural Science Foundation of China (NSFC), Grant No.~11973019, and Fudan University, Grant No.~IDH1512060. A.B.A. also acknowledges the support from the Shanghai Government Scholarship (SGS). S.N. acknowledges support from the Excellence Initiative at Eberhard-Karls Universit\"at T\"ubingen and from the Alexander von Humboldt Foundation.

%%%%%%%%%%%%%%%%%%%%%%%%%%%%%%%


\begin{thebibliography}{99}

\bibitem{kerr} 
  R.~P.~Kerr,
  %``Gravitational field of a spinning mass as an example of algebraically special metrics,''
  Phys.\ Rev.\ Lett.\  {\bf 11}, 237 (1963). 

\bibitem{k1} 
  R.~H.~Price,
  %``Nonspherical perturbations of relativistic gravitational collapse. 1. Scalar and gravitational perturbations,''
  Phys.\ Rev.\ D {\bf 5}, 2419 (1972).

\bibitem{k2} 
  C.~Bambi,
  %``Astrophysical Black Holes: A Compact Pedagogical Review,''
  Annalen Phys.\  {\bf 530}, 1700430 (2018)
  [arXiv:1711.10256 [gr-qc]].

\bibitem{k3} 
  C.~Bambi, D.~Malafarina and N.~Tsukamoto,
  %``Note on the effect of a massive accretion disk in the measurements of black hole spins,''
  Phys.\ Rev.\ D {\bf 89}, 127302 (2014)
  [arXiv:1406.2181 [gr-qc]].

\bibitem{k4} 
  C.~Bambi, A.~D.~Dolgov and A.~A.~Petrov,
  %``Black holes as antimatter factories,''
  JCAP {\bf 0909}, 013 (2009)
  [arXiv:0806.3440 [astro-ph]].
  
\bibitem{valtonen} 
  M.~J.~Valtonen {\it et al.},
  %``Measuring the spin of the primary black hole in OJ287,''
  Astrophys.\ J.\  {\bf 709}, 725 (2010)
  [arXiv:0912.1209 [astro-ph.HE]].
    
\bibitem{ligo} 
  B.~P.~Abbott {\it et al.} [LIGO Scientific and Virgo Collaborations],
  %``Observation of Gravitational Waves from a Binary Black Hole Merger,''
  Phys.\ Rev.\ Lett.\  {\bf 116}, 061102 (2016)
  [arXiv:1602.03837 [gr-qc]].  
  
\bibitem{yyp} 
  N.~Yunes, K.~Yagi and F.~Pretorius,
  %``Theoretical Physics Implications of the Binary Black-Hole Mergers GW150914 and GW151226,''
  Phys.\ Rev.\ D {\bf 94}, 084002 (2016)
  [arXiv:1603.08955 [gr-qc]].   
  
\bibitem{noi} 
  A.~Tripathi, S.~Nampalliwar, A.~B.~Abdikamalov, D.~Ayzenberg, C.~Bambi, T.~Dauser, J.~A.~Garcia and A.~Marinucci,
  %``Toward Precision Tests of General Relativity with Black Hole X-Ray Reflection Spectroscopy,''
  Astrophys.\ J.\  {\bf 875}, 56 (2019)
  [arXiv:1811.08148 [gr-qc]].  
  
\bibitem{eht} 
  K.~Akiyama {\it et al.} [Event Horizon Telescope Collaboration],
  %``First M87 Event Horizon Telescope Results. I. The Shadow of the Supermassive Black Hole,''
  Astrophys.\ J.\  {\bf 875}, L1 (2019)
  [arXiv:1906.11238 [astro-ph.GA]].  
  
\bibitem{ghez} 
  T.~Do {\it et al.},
  %``Relativistic redshift of the star S0-2 orbiting the Galactic center supermassive black hole,''
  Science {\bf 365}, 664 (2019)
  [arXiv:1907.10731 [astro-ph.GA]].  
  
\bibitem{m1} 
  S.~Capozziello and M.~De Laurentis,
  %``Extended Theories of Gravity,''
  Phys.\ Rept.\  {\bf 509}, 167 (2011)
  [arXiv:1108.6266 [gr-qc]].

\bibitem{m2} 
  C.~A.~R.~Herdeiro and E.~Radu,
  %``Kerr black holes with scalar hair,''
  Phys.\ Rev.\ Lett.\  {\bf 112}, 221101 (2014)
  [arXiv:1403.2757 [gr-qc]].  
  
\bibitem{n1} 
  G.~Dvali and C.~Gomez,
  %``Black Hole's Quantum N-Portrait,''
  Fortsch.\ Phys.\  {\bf 61}, 742 (2013)
  [arXiv:1112.3359 [hep-th]].  
  
\bibitem{n2} 
  S.~B.~Giddings,
  %``Astronomical tests for quantum black hole structure,''
  Nature Astronomy {\bf 1}, 0067 (2017)
  [arXiv:1703.03387 [gr-qc]].    

\bibitem{n3} 
  S.~B.~Giddings and D.~Psaltis,
  %``Event Horizon Telescope Observations as Probes for Quantum Structure of Astrophysical Black Holes,''
  Phys.\ Rev.\ D {\bf 97}, 084035 (2018)
  [arXiv:1606.07814 [astro-ph.HE]]. 
  
\bibitem{n4} 
  R.~Carballo-Rubio, F.~Di Filippo, S.~Liberati and M.~Visser,
  arXiv:1908.03261 [gr-qc].    
  
 \bibitem{em1} 
  C.~Bambi,
  %``Testing black hole candidates with electromagnetic radiation,''
  Rev.\ Mod.\ Phys.\  {\bf 89}, 025001 (2017)
  [arXiv:1509.03884 [gr-qc]].
 
\bibitem{em2} 
  T.~Johannsen,
  %``Sgr A* and General Relativity,''
  Class.\ Quant.\ Grav.\  {\bf 33}, 113001 (2016)
  [arXiv:1512.03818 [astro-ph.GA]].
 
\bibitem{em3} 
  J.~Schee and Z.~Stuchlik,
  %``Optical phenomena in the field of braneworld Kerr black holes,''
  Int.\ J.\ Mod.\ Phys.\ D {\bf 18}, 983 (2009)
  [arXiv:0810.4445 [astro-ph]]. 
 
\bibitem{em4} 
  C.~Bambi and K.~Freese,
  %``Apparent shape of super-spinning black holes,''
  Phys.\ Rev.\ D {\bf 79}, 043002 (2009)
  [arXiv:0812.1328 [astro-ph]]. 
   
\bibitem{em5} 
  J.~Schee and Z.~Stuchlik,
  %``Profiles of emission lines generated by rings orbiting braneworld Kerr black holes,''
  Gen.\ Rel.\ Grav.\  {\bf 41}, 1795 (2009)
  [arXiv:0812.3017 [astro-ph]]. 

\bibitem{em6} 
  C.~Bambi and E.~Barausse,
  %``Constraining the quadrupole moment of stellar-mass black-hole candidates with the continuum fitting method,''
  Astrophys.\ J.\  {\bf 731}, 121 (2011)
  [arXiv:1012.2007 [gr-qc]]. 
   
 \bibitem{em7} 
  C.~Bambi and L.~Modesto,
  %``Can an astrophysical black hole have a topologically non-trivial event horizon?,''
  Phys.\ Lett.\ B {\bf 706}, 13 (2011)
  [arXiv:1107.4337 [gr-qc]].  
  
\bibitem{em7b} 
  C.~Bambi,
  %``Towards the use of the most massive black hole candidates in AGN to test the Kerr paradigm,''
  Phys.\ Rev.\ D {\bf 85}, 043001 (2012)
  [arXiv:1112.4663 [gr-qc]].  
  
\bibitem{em8} 
  T.~Johannsen and D.~Psaltis,
  %``Testing the No-Hair Theorem with Observations in the Electromagnetic Spectrum. IV. Relativistically Broadened Iron Lines,''
  Astrophys.\ J.\  {\bf 773}, 57 (2013)
  [arXiv:1202.6069 [astro-ph.HE]].  
  
\bibitem{em8b} 
  H.~Krawczynski,
  %``Tests of General Relativity in the Strong Gravity Regime Based on X-Ray Spectropolarimetric Observations of Black Holes in X-Ray Binaries,''
  Astrophys.\ J.\  {\bf 754}, 133 (2012)
  [arXiv:1205.7063 [gr-qc]].  
   
\bibitem{em8c} 
  C.~Bambi,
  %``A code to compute the emission of thin accretion disks in non-Kerr space-times and test the nature of black hole candidates,''
  Astrophys.\ J.\  {\bf 761}, 174 (2012)
  [arXiv:1210.5679 [gr-qc]].   
   
\bibitem{em9} 
  L.~Kong, Z.~Li and C.~Bambi,
  %``Constraints on the spacetime geometry around 10 stellar-mass black hole candidates from the disk's thermal spectrum,''
  Astrophys.\ J.\  {\bf 797}, 78 (2014)
  [arXiv:1405.1508 [gr-qc]].
     
\bibitem{em10} 
  Z.~Li and C.~Bambi,
  %``Distinguishing black holes and wormholes with orbiting hot spots,''
  Phys.\ Rev.\ D {\bf 90}, 024071 (2014)
  [arXiv:1405.1883 [gr-qc]]. 
  
\bibitem{em11} 
  C.~Bambi, J.~Jiang and J.~F.~Steiner,
  %``Testing the no-hair theorem with the continuum-fitting and the iron line methods: a short review,''
  Class.\ Quant.\ Grav.\  {\bf 33}, 064001 (2016)
  [arXiv:1511.07587 [gr-qc]].  
  
\bibitem{will} 
  C.~M.~Will,
  %``The Confrontation between General Relativity and Experiment,''
  Living Rev.\ Rel.\  {\bf 17}, 4 (2014)
  [arXiv:1403.7377 [gr-qc]].  
  
\bibitem{gw1} 
  J.~R.~Gair, M.~Vallisneri, S.~L.~Larson and J.~G.~Baker,
  %``Testing General Relativity with Low-Frequency, Space-Based Gravitational-Wave Detectors,''
  Living Rev.\ Rel.\  {\bf 16}, 7 (2013)
  [arXiv:1212.5575 [gr-qc]].  
  
\bibitem{gw2} 
  N.~Yunes and X.~Siemens,
  %``Gravitational-Wave Tests of General Relativity with Ground-Based Detectors and Pulsar Timing-Arrays,''
  Living Rev.\ Rel.\  {\bf 16}, 9 (2013)
  [arXiv:1304.3473 [gr-qc]].  
  
\bibitem{gw3} 
  B.~P.~Abbott {\it et al.} [LIGO Scientific and Virgo Collaborations],
  %``Tests of general relativity with GW150914,''
  Phys.\ Rev.\ Lett.\  {\bf 116}, 221101 (2016)
  [Erratum: Phys.\ Rev.\ Lett.\  {\bf 121}, 129902 (2018)]
  [arXiv:1602.03841 [gr-qc]].  
  
\bibitem{gw4} 
  E.~Berti, K.~Yagi and N.~Yunes,
  %``Extreme Gravity Tests with Gravitational Waves from Compact Binary Coalescences: (I) Inspiral-Merger,''
  Gen.\ Rel.\ Grav.\  {\bf 50}, 46 (2018)
  [arXiv:1801.03208 [gr-qc]].
  
\bibitem{gw5} 
  E.~Berti, K.~Yagi, H.~Yang and N.~Yunes,
  %``Extreme Gravity Tests with Gravitational Waves from Compact Binary Coalescences: (II) Ringdown,''
  Gen.\ Rel.\ Grav.\  {\bf 50}, 49 (2018)
  [arXiv:1801.03587 [gr-qc]].   
  
\bibitem{gw6} 
  S.~B.~Giddings, S.~Koren and G.~Trevino,
  %``Exploring strong-field deviations from general relativity via gravitational waves,''
  arXiv:1904.04258 [gr-qc].  
  
\bibitem{barausse} 
  E.~Barausse and T.~P.~Sotiriou,
  %``Perturbed Kerr Black Holes can probe deviations from General Relativity,''
  Phys.\ Rev.\ Lett.\  {\bf 101}, 099001 (2008).  
  
\bibitem{r1} 
  C.~S.~Reynolds,
  %``Measuring Black Hole Spin using X-ray Reflection Spectroscopy,''
  Space Sci.\ Rev.\  {\bf 183}, 277 (2014)
  [arXiv:1302.3260 [astro-ph.HE]].  
  
\bibitem{r2} 
  Z.~Cao, S.~Nampalliwar, C.~Bambi, T.~Dauser and J.~A.~Garcia,
  %``Testing general relativity with the reflection spectrum of the supermassive black hole in 1H0707$-$495,''
  Phys.\ Rev.\ Lett.\  {\bf 120}, 051101 (2018)
  [arXiv:1709.00219 [gr-qc]].   

\bibitem{r3} 
  A.~B.~Abdikamalov {\it et al.},
  %``Testing general relativity with supermassive black holes using X-ray reflection spectroscopy,''
  MDPI Proc.\  {\bf 17}, 2 (2019)
  [arXiv:1905.08012 [gr-qc]].   

\bibitem{Kara:2014jda} 
  E.~Kara {\it et al.},
  %``Iron K and Compton hump reverberation in SWIFT J2127.4+5654 and NGC 1365 revealed by NuSTAR and XMM–Newton,''
  Mon.\ Not.\ Roy.\ Astron.\ Soc.\  {\bf 446}, 737 (2015)
  [arXiv:1410.3357 [astro-ph.HE]].

\bibitem{Kara:2019zad} 
  E.~Kara {\it et al.},
  %``The corona contracts in a black-hole transient,''
  Nature {\bf 565}, 198 (2019)
  [arXiv:1901.03877 [astro-ph.HE]]. 

\bibitem{v1} 
  C.~Bambi, A.~Cardenas-Avendano, T.~Dauser, J.~A.~Garcia and S.~Nampalliwar,
  %``Testing the Kerr black hole hypothesis using X-ray reflection spectroscopy,''
  Astrophys.\ J.\  {\bf 842}, 76 (2017)
  [arXiv:1607.00596 [gr-qc]].
    
\bibitem{v2} 
  A.~B.~Abdikamalov, D.~Ayzenberg, C.~Bambi, T.~Dauser, J.~A.~Garcia and S.~Nampalliwar,
  %``Public Release of RELXILL_NK: A Relativistic Reflection Model for Testing Einstein’s Gravity,''
  Astrophys.\ J.\  {\bf 878}, 91 (2019)
  [arXiv:1902.09665 [gr-qc]].  

\bibitem{j1} 
  J.~Garcia, T.~Dauser, C.~S.~Reynolds, T.~R.~Kallman, J.~E.~McClintock, J.~Wilms and W.~Eikmann,
  %``X-ray reflected spectra from accretion disk models. III. A complete grid of ionized reflection calculations,''
  Astrophys.\ J.\  {\bf 768}, 146 (2013)
  [arXiv:1303.2112 [astro-ph.HE]].
  
\bibitem{j2} 
  J.~Garcia {\it et al.},
  %``Improved Reflection Models of Black-Hole Accretion Disks: Treating the Angular Distribution of X-rays,''
  Astrophys.\ J.\  {\bf 782}, 76 (2014)
  [arXiv:1312.3231 [astro-ph.HE]].  
  
\bibitem{johannsen} 
  T.~Johannsen,
  %``Regular Black Hole Metric with Three Constants of Motion,''
  Phys.\ Rev.\ D {\bf 88}, 044002 (2013)
  [arXiv:1501.02809 [gr-qc]].  
 
 \bibitem{bare} 
  A.~Tripathi {\it et al.},
  %``Constraints on the Spacetime Metric around Seven “Bare” AGNs Using X-Ray Reflection Spectroscopy,''
  Astrophys.\ J.\  {\bf 874}, 135 (2019)
  [arXiv:1901.03064 [gr-qc]]. 
  
\bibitem{gs1354} 
  Y.~Xu, S.~Nampalliwar, A.~B.~Abdikamalov, D.~Ayzenberg, C.~Bambi, T.~Dauser, J.~A.~Garcia and J.~Jiang,
  %``A Study of the Strong Gravity Region of the Black Hole in GS 1354–645,''
  Astrophys.\ J.\  {\bf 865}, 134 (2018)
  [arXiv:1807.10243 [gr-qc]].   
 
\bibitem{nan} 
  N.~Lin, N.~Tsukamoto, M.~Ghasemi-Nodehi and C.~Bambi,
  %``A parametrization to test black hole candidates with the spectrum of thin disks,''
  Eur.\ Phys.\ J.\ C {\bf 75}, 599 (2015)
  [arXiv:1512.00724 [gr-qc]]. 
  
\bibitem{w13} 
  D.~J.~Walton, E.~Nardini, A.~C.~Fabian, L.~C.~Gallo and R.~C.~Reis,
  %``Suzaku observations of 'bare' active galactic nuclei,''
  Mon.\ Not.\ Roy.\ Astron.\ Soc.\  {\bf 428}, 2901 (2013)
  [arXiv:1210.4593 [astro-ph.HE]]. 

\bibitem{Jiang:2018aln} 
  J.~Jiang, D.~J.~Walton, A.~C.~Fabian and M.~L.~Parker,
  %``A Relativistic Disc Reflection Model for 1H0419-577: Multi-Epoch Spectral Analysis with XMM-Newton and NuSTAR,''
  Mon.\ Not.\ Roy.\ Astron.\ Soc.\  {\bf 483}, 2958 (2019)
  [arXiv:1811.10932 [astro-ph.HE]].  
  
\bibitem{Jiang:2019ztr} 
  J.~Jiang {\it et al.},
  %``High Density Reflection Spectroscopy – II. The density of the inner black hole accretion disc in AGN,''
  Mon.\ Not.\ Roy.\ Astron.\ Soc.\  {\bf 489}, 3436 (2019)
  [arXiv:1908.07272 [astro-ph.HE]].   
  
\bibitem{Agis-Gonzalez:2014nja} 
  B.~Agis-Gonzalez {\it et al.},
  %``Black hole spin and size of the X-ray-emitting region(s) in the Seyfert 1.5 galaxy ESO 362−G18,''
  Mon.\ Not.\ Roy.\ Astron.\ Soc.\  {\bf 443}, 2862 (2014)
  [arXiv:1407.1238 [astro-ph.HE]].  
  
\bibitem{trf} 
  C.~T.~Cunningham,
  %``The effects of redshifts and focusing on the spectrum of an accretion disk around a Kerr black hole,''
  Astrophys.\ J.\  {\bf 202}, 788 (1975).   
 
\bibitem{arnaud} 
  K.~A.~Arnaud,
  Astronomical Data Analysis Software and Systems V, {\bf 101}, 17 (1996).   
  
\bibitem{wilms} 
  J.~Wilms, A.~Allen and R.~McCray,
  %``On the Absorption of X-rays in the interstellar medium,''
  Astrophys.\ J.\  {\bf 542}, 914 (2000)
  [astro-ph/0008425].  
  
\bibitem{Willingale:2013tia} 
  R.~Willingale, R.~L.~C.~Starling, A.~P.~Beardmore, N.~R.~Tanvir and P.~T.~O'Brien,
  %``Calibration of X-ray absorption in our Galaxy,''
  Mon.\ Not.\ Roy.\ Astron.\ Soc.\  {\bf 431}, 394 (2013)
  [arXiv:1303.0843 [astro-ph.HE]].  
  
\bibitem{xillver} 
  J.~Garcia and T.~Kallman,
  %``X-ray reflected spectra from accretion disk models. I. Constant density atmospheres,''
  Astrophys.\ J.\  {\bf 718}, 695 (2010)
  [arXiv:1006.0485 [astro-ph.HE]].  
  
\bibitem{Wilkins:2014caa} 
  D.~R.~Wilkins and L.~C.~Gallo,
  %``The Comptonization of accretion disc X-ray emission: consequences for X-ray reflection and the geometry of AGN coronae,''
  Mon.\ Not.\ Roy.\ Astron.\ Soc.\  {\bf 448}, 703 (2015)
  [arXiv:1412.0015 [astro-ph.HE]].  
  
\bibitem{Ross:2005dm} 
  R.~R.~Ross and A.~C.~Fabian,
  %``A Comprehensive range of x-ray ionized reflection models,''
  Mon.\ Not.\ Roy.\ Astron.\ Soc.\  {\bf 358}, 211 (2005)
  doi:10.1111/j.1365-2966.2005.08797.x
  [astro-ph/0501116]. 
  
\bibitem{Dauser:2010ne} 
  T.~Dauser, J.~Wilms, C.~S.~Reynolds and L.~W.~Brenneman,
  %``Broad emission lines for negatively spinning black holes,''
  Mon.\ Not.\ Roy.\ Astron.\ Soc.\  {\bf 409}, 1534 (2010)
  doi:10.1111/j.1365-2966.2010.17393.x
  [arXiv:1007.4937 [astro-ph.HE]].  
  
\bibitem{inp} 
  A.~Tripathi {\it et al.}, in preparation.   
  
\bibitem{Kara:2017jdb} 
  E.~Kara, J.~A.~Garcia, A.~Lohfink, A.~C.~Fabian, C.~S.~Reynolds, F.~Tombesi and D.~R.~Wilkins,
  %``The high-Eddington NLS1 Ark 564 has the coolest corona,''
  Mon.\ Not.\ Roy.\ Astron.\ Soc.\  {\bf 468}, 3489 (2017)
  [arXiv:1703.09815 [astro-ph.HE]].   
  
\bibitem{hh} 
  H.~Liu, A.~B.~Abdikamalov, D.~Ayzenberg, C.~Bambi, T.~Dauser, J.~A.~Garcia and S.~Nampalliwar,
  %``Testing the Kerr hypothesis using X-ray reflection spectroscopy with NuSTAR data of Cygnus X-1 in the soft state,''
  Phys.\ Rev.\ D {\bf 99}, 123007 (2019)
  [arXiv:1904.08027 [gr-qc]].   
  
\bibitem{brenneman} 
  L.~W.~Brenneman and C.~S.~Reynolds,
  %``Constraining Black Hole Spin Via X-ray Spectroscopy,''
  Astrophys.\ J.\  {\bf 652}, 1028 (2006)
  [astro-ph/0608502].   
  
\bibitem{Riaz:2019bkv} 
  S.~Riaz, D.~Ayzenberg, C.~Bambi and S.~Nampalliwar,
  %``Reflection spectra of thick accretion disks,''
  Mon.\ Not.\ Roy.\ Astron.\ Soc.\  {\bf 491}, 417 (2020)
  [arXiv:1908.04969 [astro-ph.HE]]. 

\bibitem{Riaz:2019kat} 
  S.~Riaz, D.~Ayzenberg, C.~Bambi and S.~Nampalliwar,
  %``Modeling bias in supermassive black hole spin measurements,''
  arXiv:1911.06605 [astro-ph.HE]. 
  
\bibitem{yuexin} 
  Y.~Zhang, A.~B.~Abdikamalov, D.~Ayzenberg, C.~Bambi, T.~Dauser, J.~A.~Garcia and S.~Nampalliwar,
  %``About the Kerr nature of the stellar-mass black hole in GRS 1915+105,''
  Astrophys.\ J.\  {\bf 875}, 41 (2019)
  [arXiv:1901.06117 [gr-qc]].  
  
\bibitem{yuexin2} 
  Y.~Zhang, A.~B.~Abdikamalov, D.~Ayzenberg, C.~Bambi and S.~Nampalliwar,
  %``Tests of the Kerr hypothesis with GRS 1915+105 using different RELXILL flavors,''
  Astrophys.\ J.\  {\bf 884}, 147 (2019)
  [arXiv:1907.03084 [gr-qc]].  

\end{thebibliography}
\end{document}